\documentclass[twocolumn,prb,color,superscriptaddress,psfig,showpacs,amsmath,amssymb]{revtex4}

\usepackage{epsf}

\usepackage{natbib}
\setcitestyle{square,numbers}

\usepackage{tabularx} 
\usepackage{graphicx} 
\usepackage{epstopdf}
\usepackage{epsf}
\usepackage{dcolumn}
\usepackage{bm}
\usepackage{color}

\newcommand{\NaCoSbO}{Na$_{3}$Co$_{2}$SbO$_6$}
\newcommand{\RuCl}{$\alpha$-RuCl$_3$}
\newcommand{\TN}{T$_N$}
\newcommand{\LiZnSbO}{Li$_{3}$Zn$_{2}$SbO$_6$}
\newcommand{\NaCoTeO}{Na$_{2}$Co$_{2}$TeO$_6$}

\usepackage{multirow}

\usepackage[normalem]{ulem}

\usepackage{color,soul}
\setul{0.5ex}{0.5ex}
\setulcolor{red}
\def\blue{\color{blue}}

\begin{document}

\title{Magnetic phase diagram and possible Kitaev-like behavior of honeycomb-lattice antimonate \NaCoSbO\ }

\author{E. Vavilova}
\affiliation{Zavoisky Physical-Technical Institute (ZPhTI) of the Kazan Scientific Center of the Russian Academy of Sciences, 420029 Kazan, Russia}

\author{T. Vasilchikova}
\affiliation{Faculty of Physics, Moscow State University, Moscow 119991, Russia}

\author{A. Vasiliev}
\affiliation{Faculty of Physics, Moscow State University, Moscow 119991, Russia}
\affiliation{National University of Science and Technology "MISiS", Moscow 119049, Russia}
\affiliation{Department of theoretical physics and applied mathematics, Ural Federal University, Ekaterinburg 620002, Russia}

\author{D. Mikhailova}
\affiliation{Institute for Complex Materials, Leibniz Institute for Solid State and Materials Research (IFW) Dresden, 01069 Dresden, Germany}

\author{V. Nalbandyan}
\affiliation{Faculty of Chemistry, Southern Federal University, Rostov-on-Don 344090, Russia}

\author{E. Zvereva}
\thanks{deceased}
\affiliation{Faculty of Physics, Moscow State University, Moscow 119991, Russia}

\author{S.V. Streltsov}
\affiliation{ Institute of Metal Physics, Ural Branch of the Russian Academy of Sciences, Ekaterinburg 620137, Russia}
\affiliation{Department of theoretical physics and applied mathematics, Ural Federal University, Ekaterinburg 620002, Russia}

\date{\today}

\begin{abstract}
	Recent theoretical studies have suggested that Kitaev physics and such effects as formation of a mysterious spin-liquid state can be expected not only in \RuCl~and iridates, but also in conventional $3d$ transition metal compounds. Using DC and AC magnetometry, thermodynamic and $^{23}$Na nuclear magnetic resonance measurements (NMR) we studied such a candidate material \NaCoSbO. A full phase diagram of \NaCoSbO~in a wide range of magnetic fields and temperatures is presented.  The results demonstrate transformation of the antiferromagnetic structure under the external magnetic field, gradual development of the saturation phase, as well as evidence of gapped behavior in certain parts of the phase diagram.

\end{abstract}

\maketitle

\section{Introduction}\label{intro}

Since proposal of the Kitaev model realization in honeycomb \RuCl~and iridates, a lot of efforts had been concentrated on study of these systems~\cite{Takagi} The quest for the spin liquid due to the anisotropic (Kitaev) exchange interaction had been continued by considering more conventional $3d$ materials~\cite{Liu2018,Sano2018,Motome2020}. It has been known for a long time that low-energy excitations in Co$^{2+}$ can be modeled by the pseudospin \textit{j}$_{\rm eff} = 1/2$~\cite{AbragamBleaney,Khomskii2020}.  Indeed, in the octahedral surrounding, $t_{2g}$ states can be described by the effective orbital moment $l_{\rm eff} = 1$, which together with spin $s = 1/2$ leads to the total moment \textit{j}$_{\rm eff} = 1/2$. Of course, both $t_{2g}$ and $e_g$ states contribute to the exchange coupling, but in a common edge geometry (nearly 90° metal-ligand-metal bonds), the antiferromagnetic (AFM) $t_{2g} - e_g$ exchange can be largely compensated by the ferromagnetic (FM) $e_g - e_g$ terms~\cite{Khaliullin2020}, so that strongly anisotropic Ising-like exchange can dominate. This suggestion resulted in a flurry of both theoretical and experimental interest to Co$^{2+}$ compounds with a honeycomb layered geometry as a new playground for the Kitaev physics. 

However, already first \textit{ab initio} and model calculations have shown that the situation is more complicated in these systems~\cite{Das, Maksimov, Winter}. In fact, a direct exchange between $xy$ orbitals for the nearest neighbors leads to a substantial isotropic exchange, which in case of BaCo$_2$(AsO$_4$)$_2$ turned out to be an order of magnitude larger than the Kitaev term~\cite{Maksimov}. Moreover, together with this, the large exchange coupling between 3rd neighbors strongly questions possibility of Kitaev model realization in Co$^{2+}$ compounds. This explains why most of them orders at very low temperature~\cite{Lin, AgCoSbO6}. Nevertheless, an external magnetic field was suggested to stabilize an emergent state in such materials as BaCo$_2$(AsO$_4$)$_2$~\cite{Zhong} and Na$_2$Co$_2$TeO$_6$~\cite{Lin}. The nature of this field-induced state is debated now, but it appears to be related to the Kitaev physics.

In the present paper, we study honeycomb antimonate \NaCoSbO, that was recently proposed as a compound which is proximate to a Kitaev spin-liquid phase~\cite{Khaliullin2020}. The crystal structure of this material is very similar to Na$_2$Co$_2$TeO$_6$ and differs mainly in the way how Co planes are stacked: in Na$_2$Co$_2$TeO$_6$ the stacking is staggered, while in Na$_3$Co$_2$SbO$_6$ Co-honeycombs are placed directly on top of each other~\cite{Viciu2007}. Single-crystal neutron diffraction shows that at temperature 5 K Na$_3$Co$_2$SbO$_6$ orders magnetically with antiferromagnetic zigzag  structure~\cite{Yan} (note that depending on quality of samples $T_N$ varies from 4.4 to 8.3 K~\cite{Wong,Viciu2007,Yan}). Effective magnetic moment of 5.2-5.5$\mu_B$\cite{Viciu2007,Wong,Yan} strongly suggests a large orbital contribution. Rich variety of field-induced phases was observed by DC magnetometry in this material~\cite{Li}. Results of recent inelastic neutron scattering measurements are summarized in Tab.~\ref{J}. They suggest a significant~\cite{Sanders2022} or even dominating\cite{songvilay2020,Kim} Kitaev term in this material, while theoretical calculations in contrast expect large isotropic exchanges for both first and third nearest neighbors (and small Kitaev term)~\cite{Winter}.

\begin{table}[t!]
\caption{\label{J}  Available in literature exchange parameters (in meV) of Heisenberg-Kitaev model extracted from fitting of inelastic neutron scattering data. $J_1$ and $J_3$ stands for the isotropic exchange between first and third nearest neighbors (positive sign corresponds to AFM), $J_2$ was found to be negligible, $K$ is the Kitaev parameter and $\Gamma, \Gamma'$ are parameters describing coupling $x$ and $y$, and $x$ (or $y$) and $z$ components of spins (in the local coordinate system, where $z$ is directed perpendicular Co-O plaquette) for nearest neighbors. Two sets of data for \cite{Kim,Sanders2022} corresponds to two fittings with different sign of $K$.}
\begin{ruledtabular}
\begin{tabular}{llllll}
& $J_1$ & $J_3$ & $K$ & $\Gamma$  & $\Gamma'$ \\
 \hline
Ref. \cite{songvilay2020} & -2.1 & 0.8 & -9.0 & 0.3 & -0.8 \\
Ref. \cite{Kim}           & -2.1 & 1.2 & -4.0 & -0.7 & 0.6 \\
Ref. \cite{Kim}           & -4.7 & 1.0 & 3.6 & 1.3 & -1.4 \\
Ref. \cite{Sanders2022}   & -1.4 & 0.6 & -10 & -0.3 & -0.6 \\
Ref. \cite{Sanders2022}   & -5.0 & 0.6 & 2 & -4 & 0.3
\end{tabular}
\end{ruledtabular}
\end{table}

Despite the fact that in recent years a number of experimental studies of this system have appeared, the  manifestation of Kitaev physics has not been definitely confirmed or refuted here. There are several main reasons for this situation: the results were obtained in narrow individual regions of the phase diagram, and the data are strongly sample dependent. Moreover, the mutual orientation of the lattice and the external field is very important for the Kitaev behavior, but sufficiently large single crystals are still not available for comprehensive studies in a wide range of fields and temperatures. As was observed in $4d$ and $5d$ compounds, the Kitaev behavior is most pronounced in a rather narrow region of the phase diagram, where the AFM order is already suppressed by the field, and saturation has not yet been reached. In many cobalt compounds, the suppression of the AFM order occurs in relatively large fields $B >$ 8T, which complicates most of experiments, although studies by local methods (see, for example ~\cite{Kikuchi} for Na$_2$Co$_2$TeO$_6$) of the correlated dynamic region above the ordering can be the argument in a favor of possible Kitaev behavior. In \NaCoSbO\~, the suppression of the zig-zag AFM order occurs at lower fields, which gives prospects for experimental studies of the spin-liquid behavior. In order to mark a possible spin-liquid region, we carried out a step-by-step study of magnetic behavior of this material in a wide range of fields and temperatures, covering all the supposed regions of the phase diagram. Having a powder sample, we cannot definitely confirm the Kitaev behavior, but by applying the formalism used for ``classical'' Kitaev compounds, one can identify borders of the regions where such a description is possible and estimate the corresponding parameters. Therefore, these results can pave the way for further more detailed studies on single crystals, which can give unambiguously correct quantitative results.

Thus, in the present paper, using AC and DC magnetometry, heat capacity and NMR measurements, we establish a phase diagram of Na$_3$Co$_2$SbO$_6$  and investigate physical properties of this system. It is shown that magnetic moments are confined to the honeycomb plane and order below $T_{\rm N}$ = 7.6 K, while a magnetic field suppresses this ordering already at $B \approx  1.5$ T. Analysis of the experimental data suggests that the spin gap develops above this field.

The paper is organized as follows. Experimental details are summarized in Sec.~\ref{methods}.  Main results of $^{23}$Na NMR, thermodynamic and magnetometry experiments are presented in Sec.~\ref{results}.  The results are discussed and the H-T diagram is given in Sec.~\ref{pdiag}. The conclusions are summarized in Sec.~\ref{sec:concl}.  Some additional experimental data are presented in Sec.~\ref{supl}.
  
\section{Experimental details}\label{methods}
\begin{figure}[t!]
	\includegraphics[width=0.95\columnwidth]{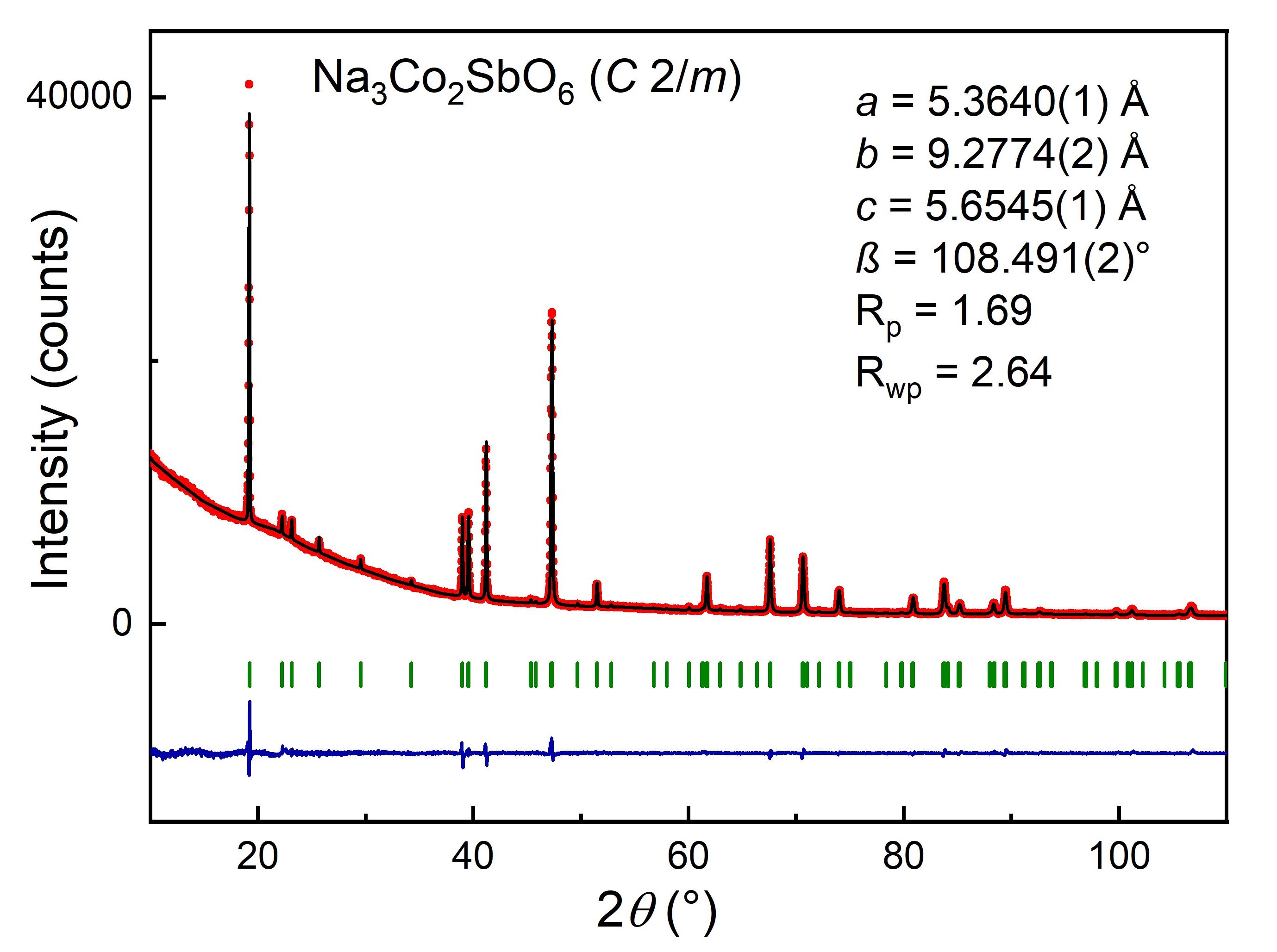}
	\caption{ (Color online) X-ray powder diffraction pattern of \NaCoSbO~with the observed (red dots) and calculated (black solid line) curves together with their difference curves (blue solid line) and Bragg positions given as green vertical lines. } 
	\label{Xray}
\end{figure}
\begin{table}[b!]
	\renewcommand{\arraystretch}{1.5}
	\caption{\label{Table} Crystal lattice parameters of \NaCoSbO. Present results are given in the last line.}
	\begin{center}
		\begin{tabular}{lcccc}
			\hline
			\hline
			Radiation  & $a$, \AA & $b$, \AA & $c$, \AA & $\beta$, $^{\circ}$\\
			\cline{1-5}
			 Neutron~\cite{Wong} & 5.35648(10) & 9.28723(19) & 5.65100(7) & 108.358(1)\\
			 Synchrotron~\cite{Stratan} &	5.3649(6) &	9.2782(3) &	5.6533(5) &	108.490(8)\\
			Co K$_{\alpha1}$ & 5.3640(1) & 9.2774(2) & 5.6545(1) & 108.491(2)\\
			\hline
			\hline
		\end{tabular}
	\end{center}
\end{table}

The polycrystalline \NaCoSbO\ sample was prepared by a conventional two-step solid-state synthesis. First, stoichiometric amounts of Sb$_2$O$_3$ (99.6\%, Alfa Aesar), Co$_3$O$_4$ (99.9\%, Sigma Aldrich) and 5\% excess of Na$_2$CO$_3$ (99.9\%, Merck) were ground together and annealed at 800°C in air for 15 h. After cooling down to room temperature, the mixture was ground again, pressed into a pellet and annealed in air at 980°C for 40 h followed by quenching in air.  After synthesis, the sample was stored in an Ar-filled glove-box with the H$_2$O/O$_2$ contents less than 0.1 ppm. The phase purity of the sample was verified by powder X-ray diffraction, using STOE STADI P diffractometer with a Co-K$_{\alpha 1}$ radiation, $\lambda$ = 1.7889 \AA, in a transmission mode. Rietveld analysis using the Fullprof software~\cite{Roisnel} (see Fig.~\ref{Xray}) confirmed the structural model based on the neutron diffraction data~\cite{Wong}.

Part of the experimental data was obtained for the sample studied earlier~\cite{Stratan}. Excellent agreement of the lattice parameters (Table ~\ref{Table}), together with some thermodynamic and magnetic characteristics, proves that both samples are essentially identical.  For NMR and magnetometry experiments, the powder sample was placed in a sealed quartz container in an argon atmosphere after preparation.

Heat capacity measurements were carried out by a relaxation method using a Quantum Design PPMS system. The plate-shaped sample was obtained by cold pressing of the polycrystalline powder. Data were collected at zero magnetic field and under an applied field up to 9 T in the temperature range 2–30 K. The AC magnetic susceptibility data were collected at the magnetic field $B$ = 0.0001 T in the temperature range between 2–45 K while varying the frequency $f$ between 0.5–10 kHz. The static magnetic measurements were performed by means of a Quantum Design SQUID –magnetometer. The temperature dependence of magnetic susceptibility was measured at various magnetic fields 0 T $\leq B \leq$ 7 T in the temperature range 1.8–300 K. Nuclear magnetic resonance (NMR) measurements were performed with a Tecmag Redstone pulsed spectrometer at fixed Larmor frequencies at several external field ranges. The spectra were collected by step-by-step sweeping the field and integrating the echo signal obtained by the standard Hahn echo pulse sequence at each field step. The longitudinal relaxation rate $T_1^{–1}$ of $^{23}$Na nuclei was measured at the maximum of the spectrum magnitude with saturation recovery and stimulated echo pulse protocols. The evolution of the magnetization after the pulse sequence is described by the equation for spin 3/2~\cite{Narath}. Due to the partial overlap of the main line and quadrupole satellites in the inhomogeneously broadened powder spectra, the stretch factor $b$ was applied:  
\begin{align} 
\label{stretch}
 M \propto M_0 \left( 0.1\ {\rm exp}(-(\tau/ T_1 )^b) + 0.9\ {\rm exp}(-(6\tau/ T_1 )^b) \right),
\end{align}
where $\tau$ is a delay between the pulses and T$_1$ is a spin-lattice relaxation time.

\section{Experimental results and discussion}\label{results}

\subsection{Specific heat}\label{cp}

\begin{figure*}[htb]
	\includegraphics[width=2\columnwidth]{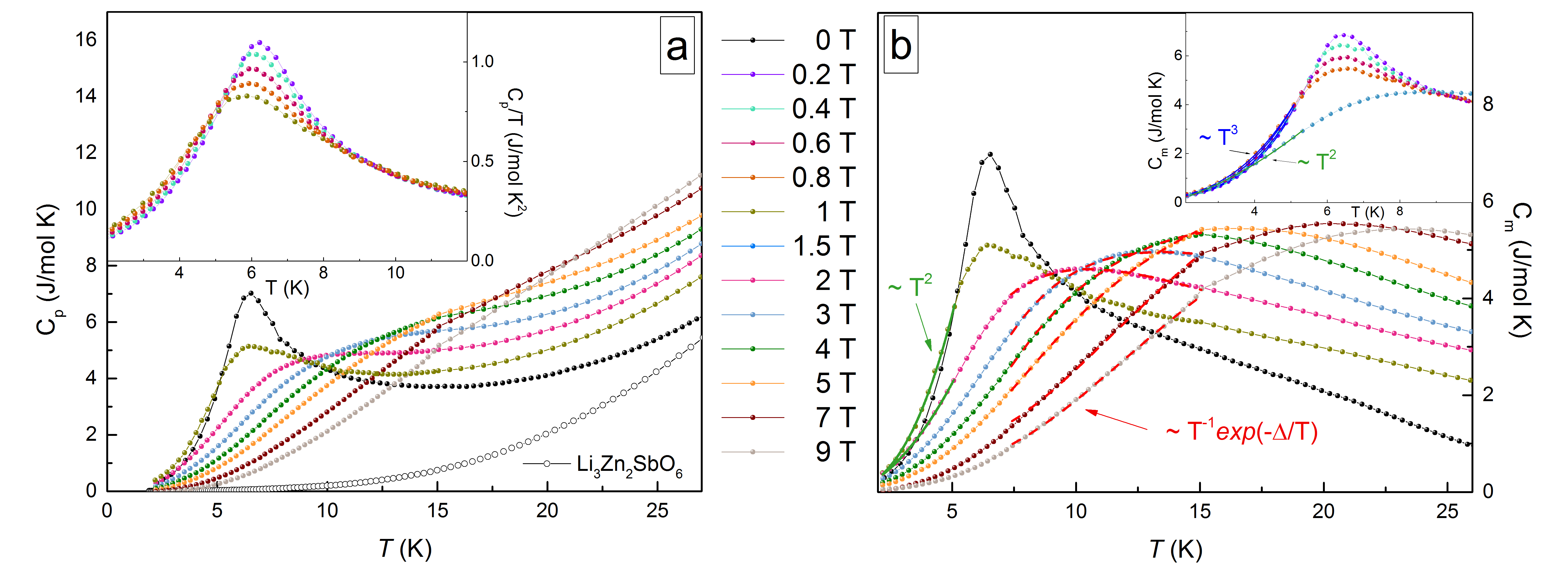}
	\caption{ (Color online) (a) The temperature dependence of the specific heat $C_p(T)$ in applied magnetic fields for \NaCoSbO~(filled circles) and non-magnetic isostructural analogue \LiZnSbO~(open circles). The inset represents $C_p/T$ at low magnetic fields. (b) The magnetic specific heat $C_m(T)$; the dashed red lines represent fit for a gapped system (Eq.~\ref{cmag2}); on the inset $C_m(T)$ at $B < 1$ T, solid blue and green lines are approximation in the frame of spin wave theory (Eq.~\ref{cmag1}).}
	\label{c}
\end{figure*}

The specific heat data at various external magnetic fields are presented in Fig.~\ref{c}. As it was found earlier~\cite{Stratan} the temperature dependence $C_p(T)$ in zero magnetic field demonstrates a distinct $\lambda$-type anomaly, associated with an onset of the magnetic order at $T_N = 6.7$ K. Under magnetic field ($B \leq$ 1 T) the singularity at $T_N$ somewhat broadens and shifts to lower temperatures (inset in Fig.~\ref{c}a). Higher fields lead to progressive weakening and eventual fading of the anomaly, which finally transforms to a broad hump; its position shifts to higher temperatures with the field (Fig. ~\ref{c}b).

In order to analyze the nature of the magnetic phase transition in \NaCoSbO~at various external fields and evaluate corresponding contribution to the specific heat we used data for its non-magnetic isostructural analogue \LiZnSbO.  The correction to this contribution for \NaCoSbO~ has been made taking into account the difference between the molar masses for each type of atom in the compound (Na-Li and Co-Zn). ~\cite{Tari}  The magnetic contribution to the specific heat was determined by subtracting the lattice contribution using the data for \LiZnSbO\ (Fig.~\ref{c}b). Analysis of the low-temperature part of $C_m(T)$ was performed in the terms of the spin-wave (SW) theory~\cite{deJongh}  
\begin{equation} 
\label{cmag1}
C_{SW} \propto T^{d/n},
\end{equation}
where $C{\rm _{SW}}$ is low-temperature specific heat due to spin-wave excitations, $d$ stands for the dimensionality of the magnetic lattice and $n$ is defined as the exponent in the dispersion relation $\omega(q) \sim q^n$. The analysis has shown that below $T_N$ the C$_m$(T) dependence follows satisfactorily $T^3$ – law at low fields $B <$ 1 T (see inset in Fig.~\ref{MT}b). This result implies a presence of antiferromagnetic magnons consistent with 3D antiferromagnetic ordering at the low temperatures in this field range. The same procedure for fields 1 T $< B \leq$ 2 T reveals $d$ = 2, indicating the predominance of 2D AFM exchanges in the studied compound.  It is noteworthy that $B \approx\ $1 T corresponds to the metamagnetic transition observed in ~\cite{Stratan, Wong}. In the vicinity of quantum critical point, temperature dependence of magnetic specific heat obeys power law $C_m$ $\sim$ $T^2$~\cite{Wolter}.   It is worth mentioning that the low-temperature magnetic specific heat in \NaCoSbO~for fields $B = 2$ T and higher can be described by the same model with the spin gap as was used for \RuCl~\cite{Wolter}
\begin{equation} 
C_{m} \propto \frac{{\rm exp}(-\Delta/k_BT)}{T}.
\label{cmag2}
\end{equation}
In order to estimate the energy gap, $C_m(T)$ was fitted by ~\eqref{cmag1}, the results are shown by dashed curves on Fig.~\ref{c}b.

\subsection{Static magnetization}\label{DC}

\begin{figure}[htb]
	\includegraphics[width=1\columnwidth]{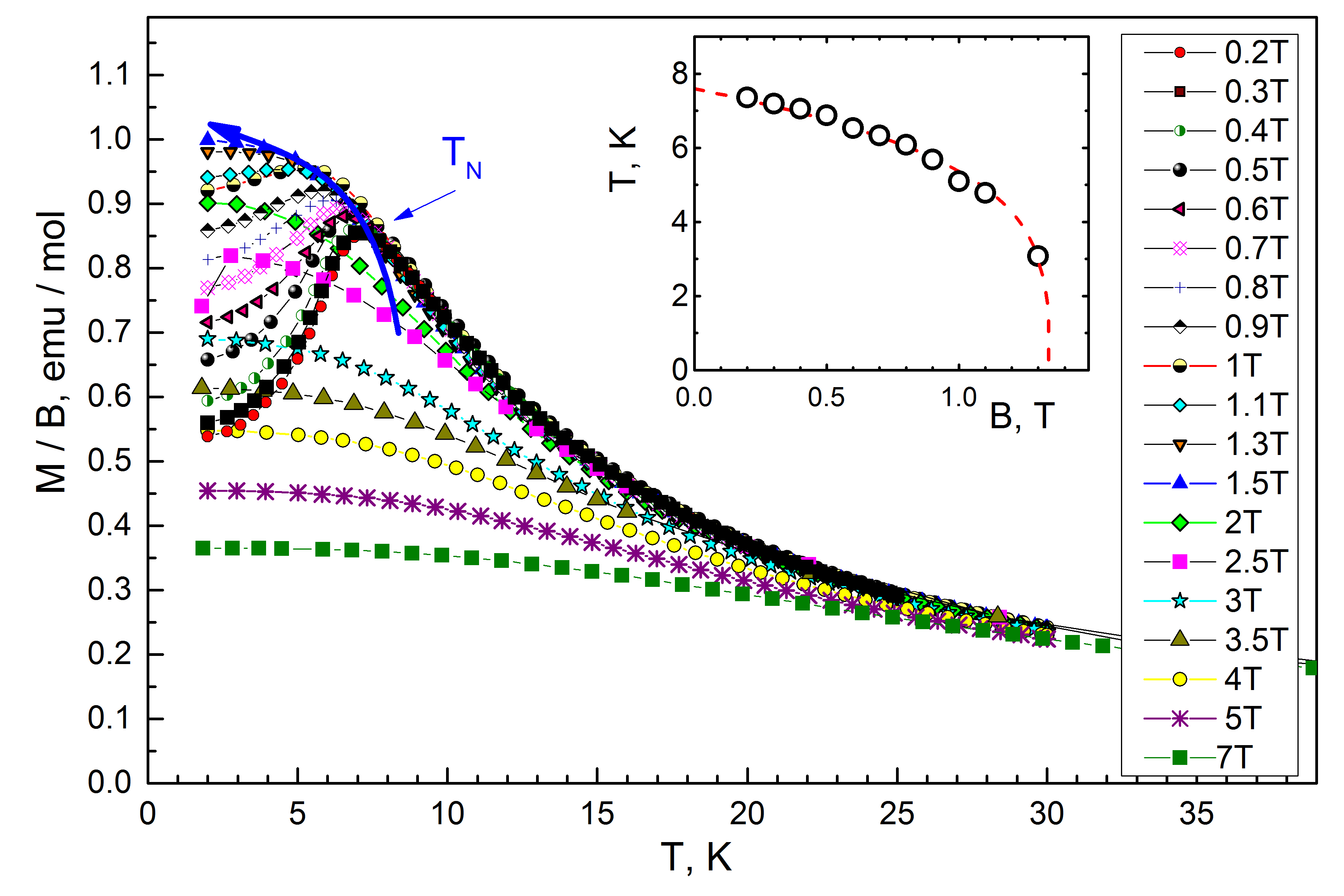}
	\caption{ (Color online)  Temperature dependence of magnetic susceptibility $\chi = M/B$ at various external fields for \NaCoSbO~.   Insert: Field dependence of the N\'eel temperature fitted by critical scenario (see text).}
 \label{MT}
\end{figure}

In accordance with previously reported data~\cite{Stratan}, the static susceptibility $\chi = M/B$ of \NaCoSbO\ (Fig.~\ref{MT}) demonstrates an antiferromagnetic behavior with low-field Weiss temperature $\Theta \approx -10$ K. For low external fields $B <$ 1.5 T, magnetic susceptibility  passes through a maximum with decrease in temperature and then drops. From the maximum $\chi (T)$, one can find the N\'eel temperature $T_N$ = 7.6 K at 0.05 T and it decreases with increase in magnetic field. The fit of  field dependence of \TN\ (Fig.~\ref{MT} insert) by expression $T_N \sim (1 - B/B_C )^{\textit{z}\nu}$ with \textit{z}$\nu$ = 0.25 $\pm$ 0.013 yields  the field of the suppression of 3D AFM order $B_C = 1.33 \pm 0.002$ T. The critical exponent \textit{z}$\nu$ is consistent with the magnetization and neutron scattering data for \RuCl~\cite{Sears}.
At the same time, 
the NMR data (see below) indicate some features typical for the AFM transition even above 1.3 T. Apparently, the strong anisotropy of the magnetic properties entails a difference in the field of AFM order suppression for different orientations of the powder crystallites.  The M/H(T) data for some directions of single crystals~\cite{Li, Yan} show that the magnitude of the peak at a field parallel to the honeycomb plane is much larger than at perpendicular direction. Therefore, the Néel temperature in the powder sample is determined mainly by crystallites of this orientation. In addition, there is a strong anisotropy even in the plane ~\cite{Li}. Thus, below we will assume that the AFM order exists in the fields below $B^* \approx$ 2 T. The $M/B (T )$ data obtained at $B \geqslant 2.7$ T indicate the saturation developed in the studied sample.  

\subsection{Nuclear magnetic resonance}~\label{NMR}

\begin{figure}[t!]
	\includegraphics[width=1\columnwidth]{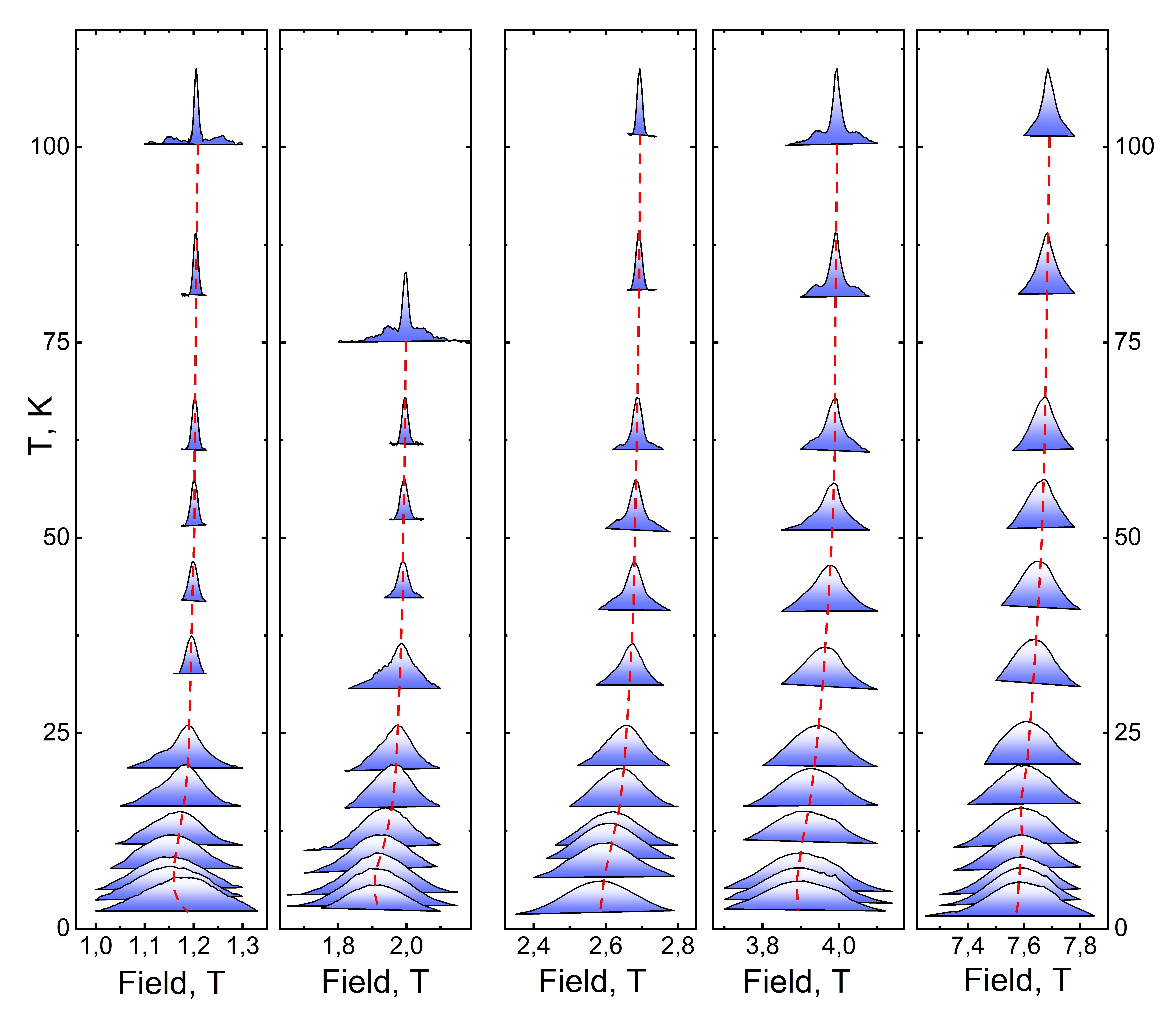}
	\caption{$^{23}$Na powder NMR spectra for different frequencies and field ranges.} 
	\label{spectra}
\end{figure}

To study the local properties of the magnetic spin system, we performed NMR measurements on the powder sample of \NaCoSbO\ on the $^{23}$Na nuclei which have a spin $I = 3/2$, the gyromagnetic ratio $\gamma_n$ = 11.2685 MHz/T, and the quadruple moment Q = 0.104 10$^{-28}$ m$^2$. The measurements were carried out at several frequencies 10.7, 13.512, 22.52, 30.33, 45.04 and 86.714 MHz in the corresponding field ranges since the effect of an external field on local magnetic characteristics is of particular interest. Temperature transformation of NMR spectra in some of field ranges is presented in Fig.~\ref{spectra}.

At temperatures of the order of 100 K and above, a classical powder spectrum with quadrupole shoulders can be recorded. As the temperature decreases, the line broadens and shifts towards lower fields. The low temperature spectrum at $B < B^*$ has a flat top but does not have a stepwise rectangular shape that is typical for antiferromagnets~\cite{Yamada}. This can be understood assuming the fact that the total width of the spectrum at low T is comparable to the quadrupole splitting and the overlap of the main line and quadrupole satellites smears the shape of the spectrum  resulting in a bell-like profile. The same reason makes the modelling of the line powder profile technically impossible and one cannot resolve the positions of the spectral contributions from the parallel and perpendicular orientations of the crystallites.

\begin{figure*}[htb]
	\includegraphics[width=1.9\columnwidth]{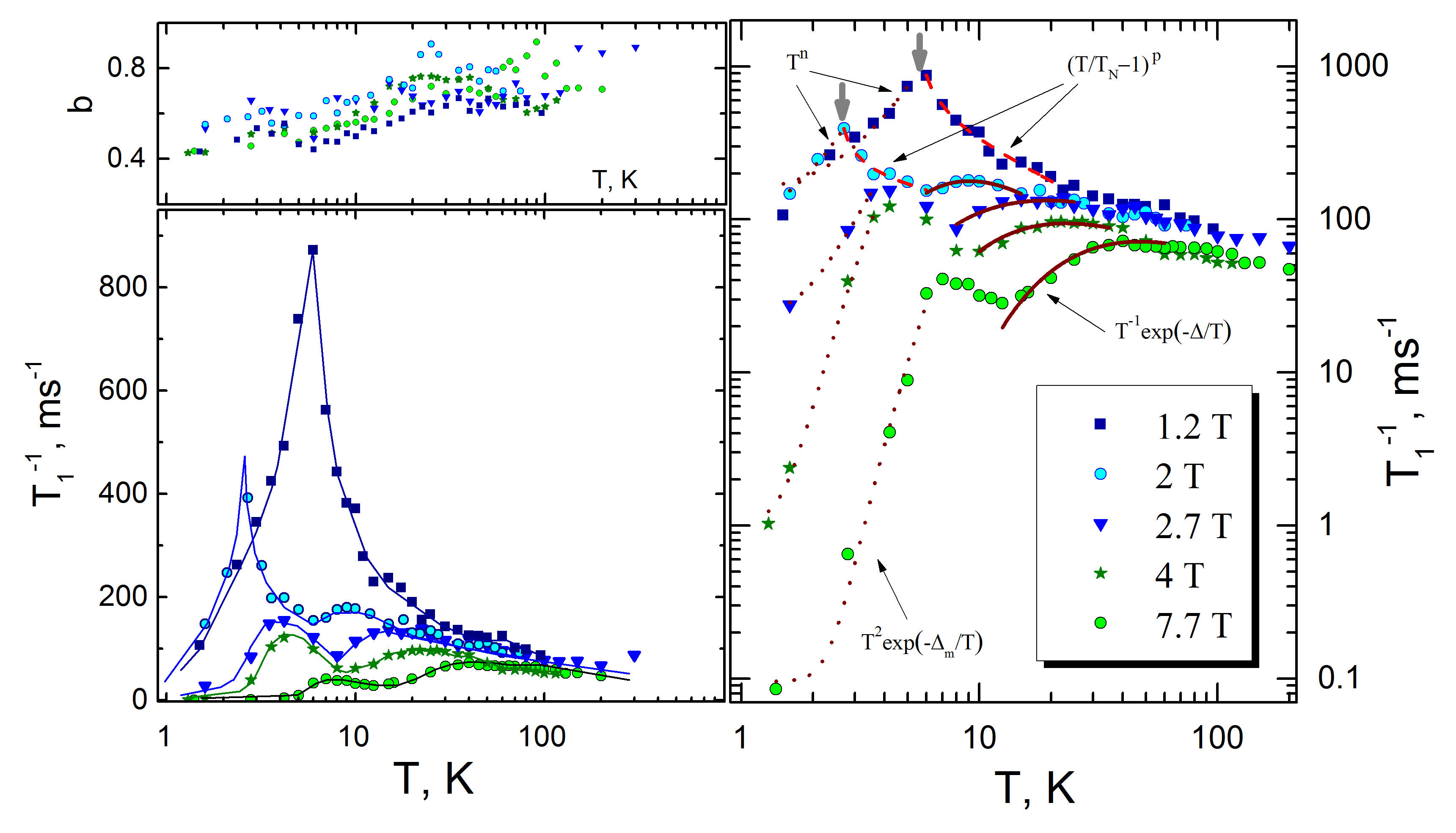}
	\caption{Left panel: Temperature dependence of $^{23}$Na spin-lattice relaxation obtained in different external fields.   Lines are guides for eye. Upper panel presents stretched coefficients from equation~\ref{stretch}. Right panel: $T_1^{-1}(T)$  in log-log scale. Solid brown lines present fit to equation $T_1^{-1} \sim T^{-n}\exp(-\Delta/T)$, where $n = 1$. Red dashed lines describe a critical behavior of the relaxation (equation~\ref{NMR1}). Brown dotted lines correspond to equation \eqref{Deltam} with $n=2$ for $B > 2.5$ T and $T_1^{-1} \sim T^{n}$ for $B < 2.5$ T. \blue {Bold grey arrows mark the transition temperatures to AFM ordered state. }}   
\label{relax}	
\end{figure*}

The dynamics of the electron spin system determines the process of nuclear spin-lattice relaxation  $T_1^{-1}$ (see Fig. ~\ref{relax}, left panels). The relaxation rate at external fields $B \leq B^*$ exhibits a sharp peak at $T < 6$ K indicating the magnetic phase transition \TN . The temperature of this peak decreases with increasing field, similarly to the N\'eel temperature determined from the static magnetic susceptibility data. Relaxation in the nearest upper vicinity of \TN~can be described in terms of the critical exponent
\begin {equation} 
(T_1^{-1})⁄(T_{1\infty}^{-1} ) \propto ((T-T_N)⁄T_N )^{-\textit{p}},
\label{NMR1}
\end{equation}
where $\textit{p} = \nu(\textit{z}-\eta)$ ~\cite{Borsa}. The obtained critical parameters are \textit{p} = 0.328 $\pm$ 0.013, $T_N = 5.51 \pm 0.08$ K for 1.2 T and \textit{p} = 0.319 $\pm$ 0.031, $T_N = 2.57 \pm 0.07$ K for 2 T. Such a small value of \textit{p} is typical for XY-triangular and honeycomb lattices~\cite{Kurbakov, Itoh} and can be estimated as close to the calculated \textit{p} = 0.32 obtained assuming $\nu \approx$ 0.5~\cite{Plumer} and \textit{z} = 0.64~\cite{Zhang}. It confirms the results of paragraph ~\ref{cp}: in-plane 2D spin correlations at $B >$ 1 T play a key role in the magnetism of \NaCoSbO~including onset of a three-dimensional AFM order. 

With a further increase of the field, a less pronounced but clearly observed maximum appears on the temperature dependence of the relaxation rate.  Its position shifts to the higher $T$ with the field and corresponds to the plateau-like behavior of temperature dependence of static susceptibility taken in the same fields. While the electron spin fluctuations slow down with temperature, the nuclear spin-lattice relaxation depends on correlation time as ~\cite{Slichter}
\begin {equation} 
T_1^{-1} \propto \tau_c ⁄(1 + \omega_0^2 \tau_c^2),
\label{corrtime}
\end{equation}
where $\tau_c$ is the correlation time and $\omega_0$ is the Larmor frequency. Below the maximum of this dependence, $\tau_c > 1/\omega_0$. This allows attributing this maximum $T_{sat}$ to the establishment of a  quasi-static saturated phase induced by the external field. 
To characterize the spin dynamics at low temperatures, we tried to estimate the magnon gap $\Delta_m$ using a formula for a gapped three-magnon  process~\cite{Narath2} 
\begin{eqnarray}
\label{Deltam}
T_1^{-1}\propto\ T^n\exp(-\Delta_m/T),
\end{eqnarray}
where $n = 2$. At $B > B^*$, such an estimation gives a gap of the order of 4 – 16 K, which grows with the field (see Fig. ~\ref{pd}). At the same time, fitting of the $T_1^{-1}$(T) curve below 2.5 T, i.e., in the AFM state, results in an extremely low value of the gap, and the data can be described by a power law $T_1^{-1} \sim T^3$ typical for three-magnon scattering in the antiferromagnets at temperatures above the energy gap in the spin wave spectrum.  This is consistent with the small anisotropy gap value $<$ 2 K  determined from the field of metamagnetic transition  ~\cite{Stratan, Wong}. It should be noted that the number of experimental points does not allow making a reliable quantitative analysis; therefore, the presented conclusions should be only considered as an estimation. 

Above $T_N$ or the assumed saturation temperature $T_{sat}$, the temperature dependence of the relaxation rate in the high fields contains a wide hump. It is noteworthy that this hump firstly appears at 2 T, which is somewhat lower than $B^*$. This feature can be interpreted as the so-called ``low-dimensional maximum'', which characterizes the slowdown of 2D spin correlations in two-dimensional compounds. Given the possible Kitaev-Heisenberg physics in the compound under study, 
{\blue it is tempting to apply as a working hypothesis} the approach developed to describe the Kitaev behavior of nuclear spin-lattice relaxation for \RuCl ~\cite{Jansa, Nagai}. This theoretical model~\cite{Yoshitake} correlates the boundaries of this hump with the magnitude of the Kitaev exchange and propose the extended exponential function  
\begin{eqnarray}
\label{DeltaNMR}
T_1^{-1} \sim T^{-n} \exp(-\Delta/T). 
\end{eqnarray}
We evaluate the spin gap assuming $n = 1$ (see Fig.~\ref{relax}, right panel) as {\blue $\sim  8$ K at 2 T and $\sim  45$ K} at 7.7 T which is comparable to the values obtained for \RuCl. {\blue Certainly given a relatively small change of $T_1^{-1}$ in the considered temperature range the so obtained values of the gap should be considered as estimates. Nevertheless they appeared consistent with the gap values obtained from the specific heat (see the discussion below).}

\section{Discussion}~\label{pdiag}

\begin{figure*}[htb]
	\includegraphics[width=1.9\columnwidth]{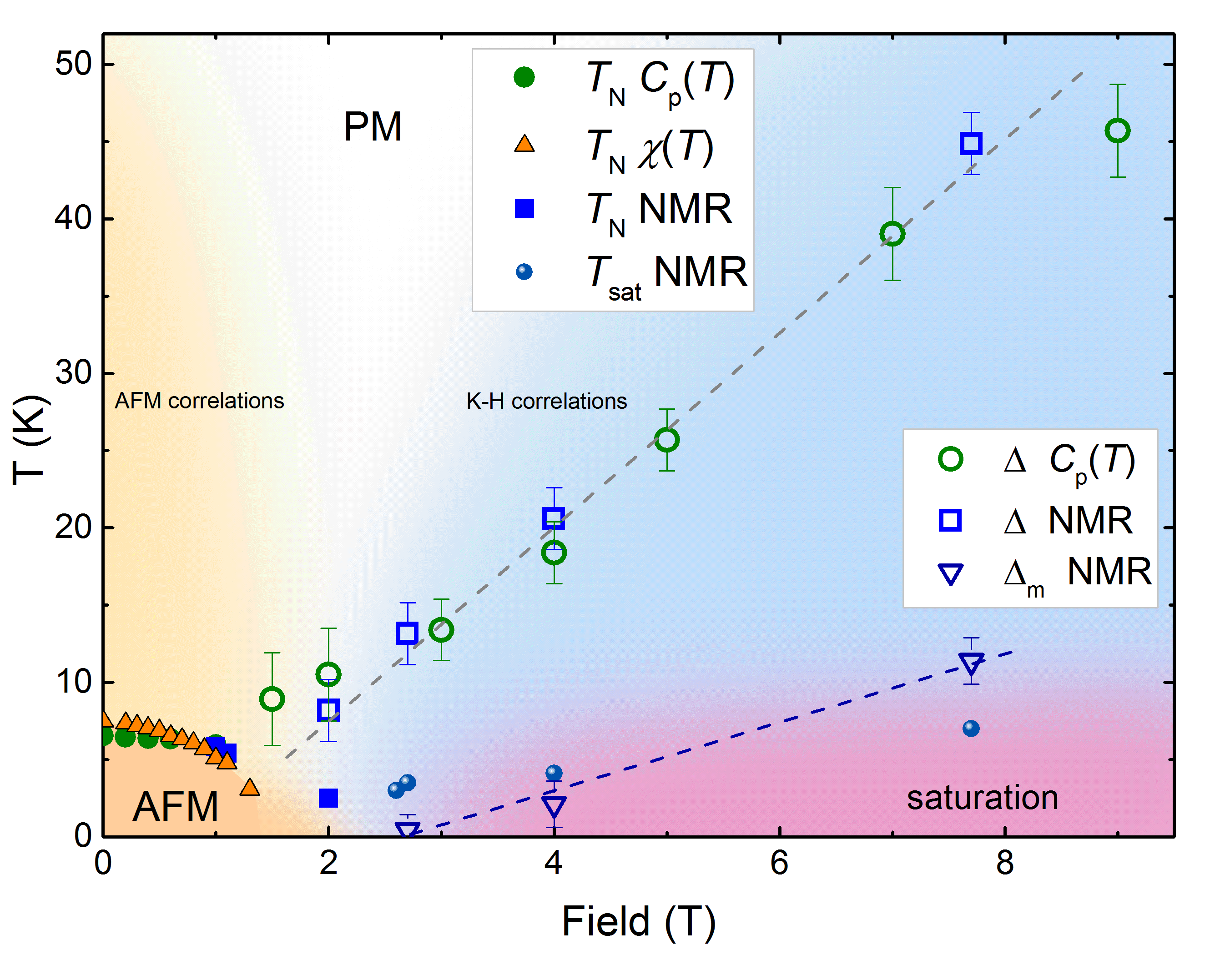}
	\caption{The magnetic phase diagram for \NaCoSbO.  $\Delta$ stands for the spin gap as defined from the specific heat using \eqref{cmag2} or by NMR measurements using \eqref{DeltaNMR}. $\Delta_m$ is the gap in magnon spectrum, if $T_1^{-1}(T)$ is fitted by \eqref{Deltam} with $n=2$}.
\label{pd}
\end{figure*}

In the T-H diagram, one can distinguish an AFM ordered phase at $B_c \leqslant 1.3$  T. Some traces of AFM order are observed up to 2 T due to different crystalline orientation in the powder sample. The AFM correlations (see Section ~\ref{supl}) are developed in a broad temperature range  above $T_N$  marked by blue color in Fig.~\ref{pd}.  

 Above $B^*$ = 2T, a wide correlated region with a spin-liquid gaped behavior is appeared (marked by blue color in Fig.~\ref{pd}). Experimentally this region is characterized by a wide hump registered by methods with very different time scales, i.e. on the temperature dependence of the specific heat, AC susceptibility, and nuclear relaxation. Such a feature is typical for low-dimensional systems. It is often associated with development of 2D short-range magnetic correlations. However, increasing the hump maximum temperature with field in \NaCoSbO~ definitely exclude the AFM origin of these correlations. 

Since \NaCoSbO~ is considered as a possible Kitaev material one can try to use a formalism developed for other Kitaev systems to describe the experimental results in this H-T region (see ~\cite{Yoshitake,Nagai,Jansa,Wolter}). This approach allows to testify to the gap character of spin excitations. Remarkably that the gap values obtained by static (specific heat, green open symbols in Fig.~\ref{pd} ) and dynamic (NMR, blue squares in Fig.~\ref{pd}) techniques are close to each other. The field behavior of the gap is almost linear. In Kitaev model~\cite{Kitaev}, the Majorana fermion gap is proportional to $B^3$, but presence of non-Kitaev interactions results in the field-induced gap for any field orientation and make it linearly dependent on the field~\cite{Song}. The gap value is comparable to the energy of the spin wave spectrum peak around $E = 1$ - $3$ meV observed by inelastic neutron scattering experiments and described in frameworks of Kitaev–Heisenberg model~\cite{Kim}. Alternative scenario of this gap development due to ferro-magnons seems to be doubtful taking into account the AC susceptibility data (see Section ~\ref{AC}).  

 The ferromagnetic excitations can be indeed observed in fields above $\approx$ 2.8 T, but at significantly lower temperatures. Here both local (NMR line shift) and bulk static susceptibility indicate the development of saturation. The characteristic growth in the relaxation rate is observed (blue circles on Fig.~\ref{pd}) , which is not so sharp as the N\'eel peak at low fields. Taking the $T_1^{-1}(T)$ dependence below this maximum, one can estimate the magnon gap $\Delta_m$. The slope of the dependence $\Delta_m/B = g\beta S/k_B$ is about 2.19 K/T (blue triangles on Fig.~\ref{pd}). This slope is consistent with the Zeeman energy of a single spin flip, i.e., a one-magnon excitation, which, with the g factor g = 3.3~\cite{Stratan} , would be 2.21 K/T.  Indeed, for the XXZ model, typical for Kitaev physics, it was shown~\cite{Wu} that the one-magnon excitations has the lowest energy in such dynamics near saturation, if the ratio of the single-ion anisotropy to $z$-component of the exchange tensor is higher than 1.33. NMR, as a local method, is sensitive to the excitations mostly on long wave vectors. Therefore, using the results of DFT calculations of the single-ion anisotropy and exchange components in \NaCoSbO ~\cite{Yan}, one can expect the observation of a one-magnon process for the saturation region.  

 It should be noted that at fields $B_c < B < B^*$ T we observed a combination of both AFM and spin-liquid features. It can be easy explained taking into account the strong anisotropy of Kitaev-Heisenberg systems behavior (see for example~\cite{Li2021}). The AFM phase field boundary is strongly depended on the crystalline particle orientation in \NaCoSbO. The data of magnetometry and magnetic neutron diffraction presented in the work ~\cite{Li} suggest that the traces of the AFM phase observed at $B_c < B < B^*$ are associated with powder particles with the a axis parallel to the field. Moreover, studies of a powder sample obviously contain a certain error in determining the magnitude and field dependence of the Kitaev gap in the spectrum of spin excitations.  Therefore, here we test applicability of the Kitaev-Heisenberg description for \NaCoSbO, whereas the exact values of spin-gap can be obtained in further measurements on single crystals.

\section{\label{sec:concl}Conclusions}
In conclusion, we performed bulk (heat capacity, DC and AC susceptibility) and local (NMR) probe measurements of the powder sample of honeycomb $j_{eff} = 1/2$ system \NaCoSbO~in a wide range of temperatures and magnetic fields.  These experiments reveal regions with a static spontaneous and field-induced magnetic order, a wide temperature range of spin correlations of various nature as well as a region of fields, where a gap in spin excitation spectrum develops. The parameters of spin excitations in different regions of the phase diagram were estimated, including the magnitude and temperature dependence of the field-induced energy gap, which determines the specific temperature dependence of bulk and local characteristics well above the saturation region. Our measurements give some evidence of possible formation of a spin-liquid-like phase  and realizing the Kitaev-Heisenberg scenario in \NaCoSbO~under external magnetic field. In this context, it would be enlightening to investigate similarly a high-quality single-crystal sample in order to clarify the details of the model parameters.

\section{Acknowledgments}

The authors are grateful to Dr. P. Maksimov, Dr. S. Winter, Dr. H.-J. Grafe, Prof. B. B\"uchner, and Dr. V. Kataev for useful discussions. A.V. acknowledges support by the Megagrant program of Russian Government through the project 075-15-2021-604. The work of D.M. was made in 2021. S.V.S. thanks Russian Science Foundation for support of theoretical analyses (RSF 20-62-46047). E.V. would like to thank financial support from the government assignment for FRC Kazan scientific Center of RAS.

\section{Appendix}~\label{supl}
\subsection{$^{23}$Na NMR line shift and hyperfine constant}~\label{shift}
\begin{figure}[t]
	\includegraphics[width=1\columnwidth]{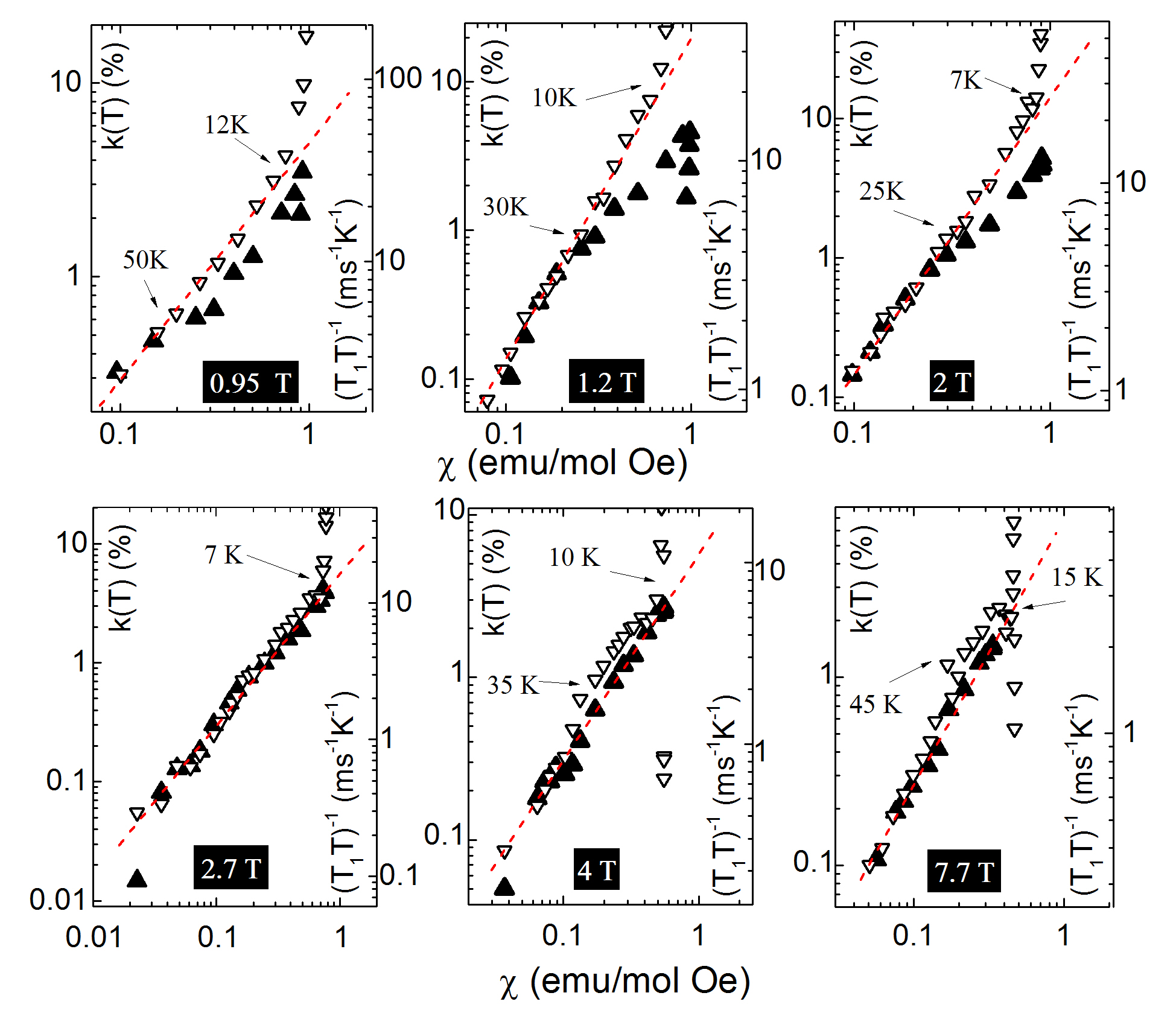}
	\caption{Dependencies of the local static $k$ (solid triangles) and dynamic $(T_1 T)^{-1}$ (open triangles) susceptibility on static bulk susceptibility at various magnetic fields.} 
\label{krelchi}
\end{figure}

The shift of the NMR line can be defined as $K$ = (($B_L - B_{res})/B_L)\times$100\%, where $B_{res}$ is the field of the spectral magnitude maximum and $B_L$ = $\omega_{Larmor}/\gamma_n$. Its temperature-dependent part corresponds to the local static susceptibility and in paramagnetic state it is proportional to the bulk susceptibility $K$ = $A\chi_{bulk}$, where A is the hyperfine tensor. The hyperfine interaction constants $A_{hf}$ taken from the linear part of the $K$ ($\chi_{bulk})$ dependencies (see Fig.~\ref{krelchi}) in each external field are found to be almost equal and the average value $A_{hf}$ is determined as 0.25$\pm $0.014  kOe/$\mu_B$. This value is comparable to that observed in \NaCoTeO ~\cite{Choi} and is significantly smaller than $A_{hf}$ for copper ~\cite{Kuo} and nickel ~\cite{Itoh} honeycomb compounds. The difference in the electronic configurations and orbital occupancy for cobalt ions seems to be responsible for the smaller orbital overlap and the transfer of spin density to sodium ions. As a result, despite the larger bulk susceptibility  in cobalt compounds compared to nickel or copper ones (see also ~\cite{Berthelot} for tellurates), the increase in the local hyperfine field on sodium nuclei is insignificant. This leads to a remarkably larger decoupling of the honeycomb planes and to a more pronounced two-dimensionality of the strongest exchanges in the cobalt compound as a result.

\subsection{local susceptibility}~\label{loc}
The NMR is well known probe of local fields, susceptibility and correlations and we use it to establish the field and temperature limits of correlated regions. The line shift  is proportional to the local field which is static on the NMR timescale $K \sim \langle H_{hf} \rangle_t \sim A_{hf} \langle S\rangle_t$. In most cases when the fluctuations of electron spins $S$ are much faster than inversed NMR frequency, time averaging can be done at zero frequency, i.e. $K \sim \chi_{loc}(0)$. In specific situations, for example in some low-dimensional systems, where the correlation time is comparable to the NMR time scale in the extended temperature range, the result of time averaging differs from that at zero frequency and  NMR shift deviates from bulk static susceptibility.

Slight deviation of $K(\chi_{bulk})$ dependence from linear at fields $B < B^*$ indicates the slowdown of cobalt spin fluctuations at low temperatures. Development of antiferromagnetic correlations with big correlation time suppressing the local field at the sodium positions located between the cobalt planes. At $B > B^*$ the local static susceptibility does not deviate from the bulk one anymore. 

\begin{figure}[t!]
	\includegraphics[width=1\columnwidth]{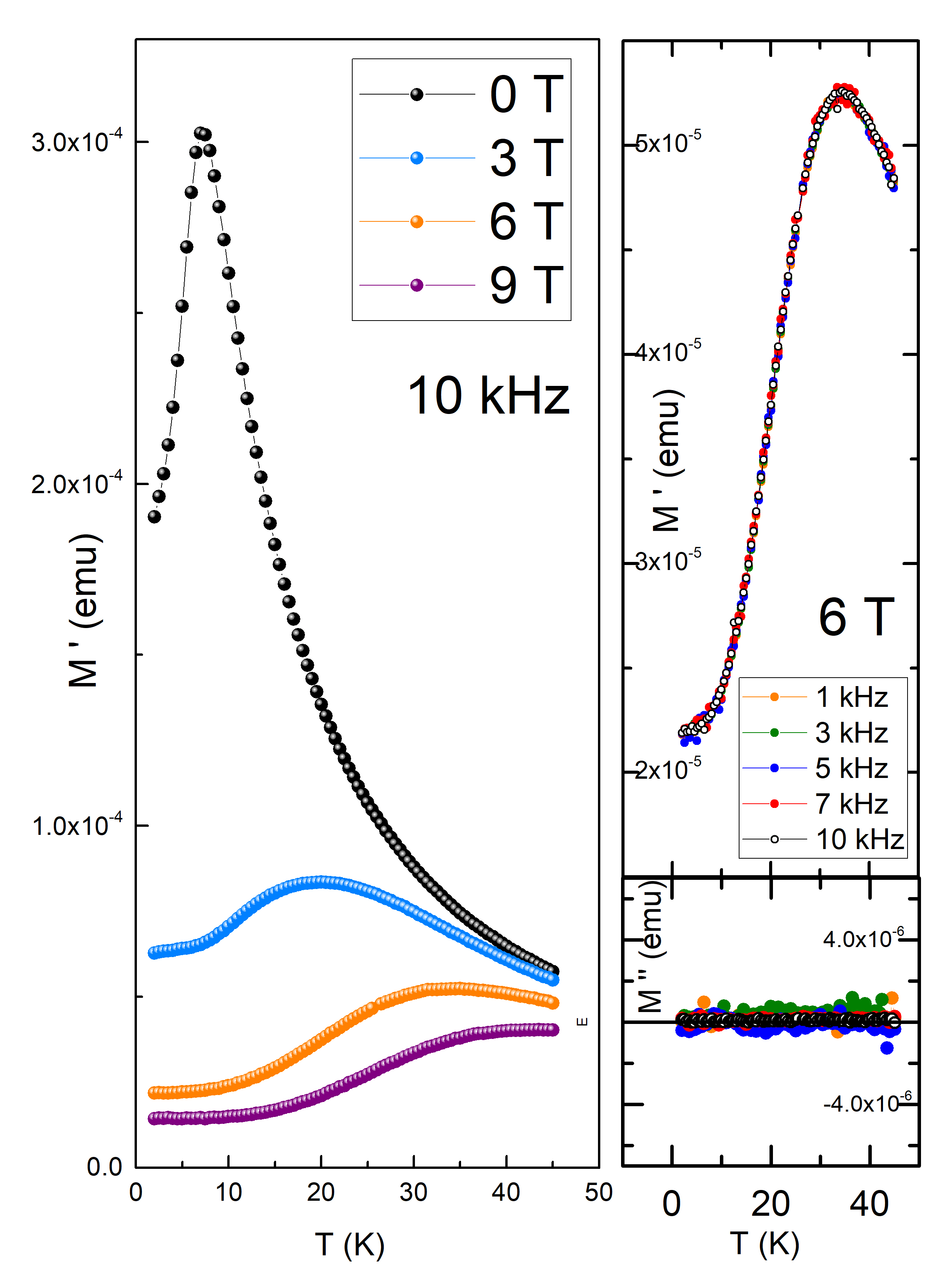}
	\caption{Left panel:  Real part $\chi$’ of the ac magnetic susceptibility at various external field. Right panel: Real part $\chi$’ (top) and imaginary part $\chi$” (bottom) of the AC magnetic susceptibility at various AC frequencies.} 
\label{ACfig}
\end{figure}

The spin-lattice relaxation can be considered as a direct probe of the local dynamic susceptibility of the electron spin subsystem~\cite{Moriya}
\begin{equation}
	\frac{1}{T_1T}\propto \sum_{\bf q}|A_{\perp}({\bf q})|^2\frac{\chi''({\bf q}, \omega_L)}{\omega_L},
\label{Moriya}
\end{equation}
where $A_{\perp}({\bf q})$ is the \textbf{q}-dependent hyperfine constant, \textbf{q} is the wave vector and $\omega_L$ is the Larmor frequency. In the paramagnetic regime $(T_1T)^{-1}$ is proportional to the static bulk susceptibility. The sharp deviation of the local dynamic susceptibility from the static bulk one occurs below 10 K in the entire range of fields (see Fig.~\ref{krelchi}), marking the critical slowing down of the electron spin fluctuations in the vicinity of the transition to the static state. In fields above $B^*$, the local dynamic susceptibility deviates from the static one already at 20 - 50 K, marking the development of dynamic correlations at a relatively low-frequency part of the spin fluctuation spectra.

\subsection{Bulk dynamic susceptibility}~\label{AC}
The AC susceptibility at zero external field exhibits a sharp peak at the N\'eel temperature. Above $B^*$, the critical low-temperature anomaly is absent, but a wide hump is observed, which shifts to higher temperatures with increasing field. At any fields, the $\chi’(T)$ curves obtained at different frequencies do not exhibit any frequency dependence (see Fig.~\ref{ACfig}). This rules out that the hump is associated with any kind of conventional glassy transition. The $\chi’$ hump is not accompanied by an anomaly of $\chi''$ and therefore one cannot attribute it to the crossover to the field-induced ferromagnetic order, since at ferro- or ferrimagnetic transition or crossover the imaginary component is present~\cite{Balanda}. On the other hand, this broad maximum shifts to higher temperatures with the applied DC field and it does not allow us to attribute it to the development of two-dimensional antiferromagnetic correlations in cobalt planes. It is remarkable that a pronounced anomaly in this temperature range is absent in the temperature dependence of both the bulk and the local static susceptibility, but it is observed in the dynamic susceptibility in both the kHz and MHz (NMR relaxation) ranges indicating the dynamic nature of the emerging magnetic state. It should be noted that the temperature of the AC susceptibility hump maximum corresponds in order of magnitude to the values of the Kitaev-Heisenberg gap obtained from nuclear relaxation analysis.

\bibliography{NaCoSbO.bib}

\begin{thebibliography}{48}
\expandafter\ifx\csname natexlab\endcsname\relax\def\natexlab#1{#1}\fi
\expandafter\ifx\csname bibnamefont\endcsname\relax
  \def\bibnamefont#1{#1}\fi
\expandafter\ifx\csname bibfnamefont\endcsname\relax
  \def\bibfnamefont#1{#1}\fi
\expandafter\ifx\csname citenamefont\endcsname\relax
  \def\citenamefont#1{#1}\fi
\expandafter\ifx\csname url\endcsname\relax
  \def\url#1{\texttt{#1}}\fi
\expandafter\ifx\csname urlprefix\endcsname\relax\def\urlprefix{URL }\fi
\providecommand{\bibinfo}[2]{#2}
\providecommand{\eprint}[2][]{\url{#2}}

\bibitem[{\citenamefont{Takagi et~al.}(2019)\citenamefont{Takagi, Takayama,
  Jackeli, Khaliullin, and Nagler}}]{Takagi}
\bibinfo{author}{\bibfnamefont{H.}~\bibnamefont{Takagi}},
  \bibinfo{author}{\bibfnamefont{T.}~\bibnamefont{Takayama}},
  \bibinfo{author}{\bibfnamefont{G.}~\bibnamefont{Jackeli}},
  \bibinfo{author}{\bibfnamefont{G.}~\bibnamefont{Khaliullin}},
  \bibnamefont{and} \bibinfo{author}{\bibfnamefont{S.~E.}
  \bibnamefont{Nagler}}, \bibinfo{journal}{Nature Reviews Physics}
  \textbf{\bibinfo{volume}{1}}, \bibinfo{pages}{264–280}
  (\bibinfo{year}{2019}),
  \urlprefix\url{https://doi.org/10.1038/s42254-019-0038-2}.

\bibitem[{\citenamefont{Liu and Khaliullin}(2018)}]{Liu2018}
\bibinfo{author}{\bibfnamefont{H.}~\bibnamefont{Liu}} \bibnamefont{and}
  \bibinfo{author}{\bibfnamefont{G.}~\bibnamefont{Khaliullin}},
  \bibinfo{journal}{Phys. Rev. B} \textbf{\bibinfo{volume}{97}},
  \bibinfo{pages}{014407} (\bibinfo{year}{2018}),
  \urlprefix\url{https://link.aps.org/doi/10.1103/PhysRevB.97.014407}.

\bibitem[{\citenamefont{Sano et~al.}(2018)\citenamefont{Sano, Kato, and
  Motome}}]{Sano2018}
\bibinfo{author}{\bibfnamefont{R.}~\bibnamefont{Sano}},
  \bibinfo{author}{\bibfnamefont{Y.}~\bibnamefont{Kato}}, \bibnamefont{and}
  \bibinfo{author}{\bibfnamefont{Y.}~\bibnamefont{Motome}},
  \bibinfo{journal}{Phys. Rev. B} \textbf{\bibinfo{volume}{97}},
  \bibinfo{pages}{014408} (\bibinfo{year}{2018}).

\bibitem[{\citenamefont{Motome et~al.}(2020)\citenamefont{Motome, Sano, Jang,
  Sugita, and Kato}}]{Motome2020}
\bibinfo{author}{\bibfnamefont{Y.}~\bibnamefont{Motome}},
  \bibinfo{author}{\bibfnamefont{R.}~\bibnamefont{Sano}},
  \bibinfo{author}{\bibfnamefont{S.}~\bibnamefont{Jang}},
  \bibinfo{author}{\bibfnamefont{Y.}~\bibnamefont{Sugita}}, \bibnamefont{and}
  \bibinfo{author}{\bibfnamefont{Y.}~\bibnamefont{Kato}}, \bibinfo{journal}{J.
  Phys. Cond. Matt.} \textbf{\bibinfo{volume}{32}}, \bibinfo{pages}{404001}
  (\bibinfo{year}{2020}), ISSN \bibinfo{issn}{23318422}.

\bibitem[{\citenamefont{Abragam and Bleaney}(1970)}]{AbragamBleaney}
\bibinfo{author}{\bibfnamefont{A.}~\bibnamefont{Abragam}} \bibnamefont{and}
  \bibinfo{author}{\bibfnamefont{B.}~\bibnamefont{Bleaney}},
  \emph{\bibinfo{title}{Electron Paramagnetic Resonance of Transition Ions}}
  (\bibinfo{publisher}{Clarendon Press, Oxford}, \bibinfo{year}{1970}).

\bibitem[{\citenamefont{Khomskii and Streltsov}(2021)}]{Khomskii2020}
\bibinfo{author}{\bibfnamefont{D.~I.} \bibnamefont{Khomskii}} \bibnamefont{and}
  \bibinfo{author}{\bibfnamefont{S.~V.} \bibnamefont{Streltsov}},
  \bibinfo{journal}{Chem. Rev.} \textbf{\bibinfo{volume}{121}},
  \bibinfo{pages}{2992} (\bibinfo{year}{2021}).

\bibitem[{\citenamefont{Liu et~al.}(2020)\citenamefont{Liu, Chaloupka, and
  Khaliullin}}]{Khaliullin2020}
\bibinfo{author}{\bibfnamefont{H.}~\bibnamefont{Liu}},
  \bibinfo{author}{\bibfnamefont{J.}~\bibnamefont{Chaloupka}},
  \bibnamefont{and}
  \bibinfo{author}{\bibfnamefont{G.}~\bibnamefont{Khaliullin}},
  \bibinfo{journal}{Phys. Rev. Lett.} \textbf{\bibinfo{volume}{125}},
  \bibinfo{pages}{047201} (\bibinfo{year}{2020}),
  \urlprefix\url{https://link.aps.org/doi/10.1103/PhysRevLett.125.047201}.

\bibitem[{\citenamefont{Das et~al.}(2021)\citenamefont{Das, Voleti,
  Saha-Dasgupta, and Paramekanti}}]{Das}
\bibinfo{author}{\bibfnamefont{S.}~\bibnamefont{Das}},
  \bibinfo{author}{\bibfnamefont{S.}~\bibnamefont{Voleti}},
  \bibinfo{author}{\bibfnamefont{T.}~\bibnamefont{Saha-Dasgupta}},
  \bibnamefont{and}
  \bibinfo{author}{\bibfnamefont{A.}~\bibnamefont{Paramekanti}},
  \bibinfo{journal}{Phys. Rev. B} \textbf{\bibinfo{volume}{104}},
  \bibinfo{pages}{134425} (\bibinfo{year}{2021}),
  \urlprefix\url{https://link.aps.org/doi/10.1103/PhysRevB.104.134425}.

\bibitem[{\citenamefont{Maksimov et~al.}(2022)\citenamefont{Maksimov, Ushakov,
  Pchelkina, Li, Winter, and Streltsov}}]{Maksimov}
\bibinfo{author}{\bibfnamefont{P.~A.} \bibnamefont{Maksimov}},
  \bibinfo{author}{\bibfnamefont{A.~V.} \bibnamefont{Ushakov}},
  \bibinfo{author}{\bibfnamefont{Z.~V.} \bibnamefont{Pchelkina}},
  \bibinfo{author}{\bibfnamefont{Y.}~\bibnamefont{Li}},
  \bibinfo{author}{\bibfnamefont{S.~M.} \bibnamefont{Winter}},
  \bibnamefont{and} \bibinfo{author}{\bibfnamefont{S.~V.}
  \bibnamefont{Streltsov}}, \bibinfo{journal}{Physical Review B}
  \textbf{\bibinfo{volume}{106}}, \bibinfo{pages}{165131}
  (\bibinfo{year}{2022}), \urlprefix\url{https://arxiv.org/abs/2204.09695}.

\bibitem[{\citenamefont{Winter}(2022)}]{Winter}
\bibinfo{author}{\bibfnamefont{S.~M.} \bibnamefont{Winter}},
  \bibinfo{journal}{Journal of Physics: Materials}
  \textbf{\bibinfo{volume}{5}}, \bibinfo{pages}{045003} (\bibinfo{year}{2022}),
  \urlprefix\url{https://arxiv.org/abs/2204.09856}.

\bibitem[{\citenamefont{Lin et~al.}(2021)\citenamefont{Lin, Jeong, Kim, Wang,
  Huang, Masuda, Asai, Itoh, Günther, Russina et~al.}}]{Lin}
\bibinfo{author}{\bibfnamefont{G.}~\bibnamefont{Lin}},
  \bibinfo{author}{\bibfnamefont{J.}~\bibnamefont{Jeong}},
  \bibinfo{author}{\bibfnamefont{C.}~\bibnamefont{Kim}},
  \bibinfo{author}{\bibfnamefont{Y.}~\bibnamefont{Wang}},
  \bibinfo{author}{\bibfnamefont{Q.}~\bibnamefont{Huang}},
  \bibinfo{author}{\bibfnamefont{T.}~\bibnamefont{Masuda}},
  \bibinfo{author}{\bibfnamefont{S.}~\bibnamefont{Asai}},
  \bibinfo{author}{\bibfnamefont{S.}~\bibnamefont{Itoh}},
  \bibinfo{author}{\bibfnamefont{G.}~\bibnamefont{Günther}},
  \bibinfo{author}{\bibfnamefont{M.}~\bibnamefont{Russina}},
  \bibnamefont{et~al.}, \bibinfo{journal}{Nature Communication}
  \textbf{\bibinfo{volume}{12}}, \bibinfo{pages}{5559} (\bibinfo{year}{2021}),
  \urlprefix\url{https://doi.org/10.1038/s41467-021-25567-7}.

\bibitem[{\citenamefont{Zvereva et~al.}(2016)\citenamefont{Zvereva, Stratan,
  Ushakov, Nalbandyan, Shukaev, Silhanek, Abdel-Hafiez, Streltsov, and
  Vasiliev}}]{AgCoSbO6}
\bibinfo{author}{\bibfnamefont{E.~A.} \bibnamefont{Zvereva}},
  \bibinfo{author}{\bibfnamefont{M.~I.} \bibnamefont{Stratan}},
  \bibinfo{author}{\bibfnamefont{A.~V.} \bibnamefont{Ushakov}},
  \bibinfo{author}{\bibfnamefont{V.~B.} \bibnamefont{Nalbandyan}},
  \bibinfo{author}{\bibfnamefont{I.~L.} \bibnamefont{Shukaev}},
  \bibinfo{author}{\bibfnamefont{A.~V.} \bibnamefont{Silhanek}},
  \bibinfo{author}{\bibfnamefont{M.}~\bibnamefont{Abdel-Hafiez}},
  \bibinfo{author}{\bibfnamefont{S.~V.} \bibnamefont{Streltsov}},
  \bibnamefont{and} \bibinfo{author}{\bibfnamefont{A.~N.}
  \bibnamefont{Vasiliev}}, \bibinfo{journal}{Dalton Trans.}
  \textbf{\bibinfo{volume}{45}}, \bibinfo{pages}{7373} (\bibinfo{year}{2016}),
  \urlprefix\url{http://dx.doi.org/10.1039/C6DT00516K}.

\bibitem[{\citenamefont{Zhong et~al.}(2020)\citenamefont{Zhong, Gao, Ong, and
  Cava}}]{Zhong}
\bibinfo{author}{\bibfnamefont{R.}~\bibnamefont{Zhong}},
  \bibinfo{author}{\bibfnamefont{T.}~\bibnamefont{Gao}},
  \bibinfo{author}{\bibfnamefont{N.~P.} \bibnamefont{Ong}}, \bibnamefont{and}
  \bibinfo{author}{\bibfnamefont{R.~J.} \bibnamefont{Cava}},
  \bibinfo{journal}{Science Advances} \textbf{\bibinfo{volume}{6}},
  \bibinfo{pages}{eaay6953} (\bibinfo{year}{2020}),
  \urlprefix\url{https://www.science.org/doi/abs/10.1126/sciadv.aay6953}.

\bibitem[{\citenamefont{Viciu et~al.}(2007)\citenamefont{Viciu, Huang, Morosan,
  Zandbergen, Greenbaum, Mcqueen, and Cava}}]{Viciu2007}
\bibinfo{author}{\bibfnamefont{L.}~\bibnamefont{Viciu}},
  \bibinfo{author}{\bibfnamefont{Q.}~\bibnamefont{Huang}},
  \bibinfo{author}{\bibfnamefont{E.}~\bibnamefont{Morosan}},
  \bibinfo{author}{\bibfnamefont{H.~W.} \bibnamefont{Zandbergen}},
  \bibinfo{author}{\bibfnamefont{N.~I.} \bibnamefont{Greenbaum}},
  \bibinfo{author}{\bibfnamefont{T.~M.} \bibnamefont{Mcqueen}},
  \bibnamefont{and} \bibinfo{author}{\bibfnamefont{R.}~\bibnamefont{Cava}},
  \bibinfo{journal}{Journal of Solid State Chemistry}
  \textbf{\bibinfo{volume}{180}}, \bibinfo{pages}{1060} (\bibinfo{year}{2007}).

\bibitem[{\citenamefont{Yan et~al.}(2019)\citenamefont{Yan, Okamoto, Wu, Zheng,
  Zhou, Cao, and McGuire}}]{Yan}
\bibinfo{author}{\bibfnamefont{J.-Q.} \bibnamefont{Yan}},
  \bibinfo{author}{\bibfnamefont{S.}~\bibnamefont{Okamoto}},
  \bibinfo{author}{\bibfnamefont{Y.}~\bibnamefont{Wu}},
  \bibinfo{author}{\bibfnamefont{Q.}~\bibnamefont{Zheng}},
  \bibinfo{author}{\bibfnamefont{H.~D.} \bibnamefont{Zhou}},
  \bibinfo{author}{\bibfnamefont{H.~B.} \bibnamefont{Cao}}, \bibnamefont{and}
  \bibinfo{author}{\bibfnamefont{M.~A.} \bibnamefont{McGuire}},
  \bibinfo{journal}{Phys. Rev. Materials} \textbf{\bibinfo{volume}{3}},
  \bibinfo{pages}{074405} (\bibinfo{year}{2019}),
  \urlprefix\url{https://link.aps.org/doi/10.1103/PhysRevMaterials.3.074405}.

\bibitem[{\citenamefont{Wong et~al.}(2016)\citenamefont{Wong, Avdeev, and
  Ling}}]{Wong}
\bibinfo{author}{\bibfnamefont{C.}~\bibnamefont{Wong}},
  \bibinfo{author}{\bibfnamefont{M.}~\bibnamefont{Avdeev}}, \bibnamefont{and}
  \bibinfo{author}{\bibfnamefont{C.~D.} \bibnamefont{Ling}},
  \bibinfo{journal}{Journal of Solid State Chemistry}
  \textbf{\bibinfo{volume}{243}}, \bibinfo{pages}{18} (\bibinfo{year}{2016}),
  ISSN \bibinfo{issn}{0022-4596},
  \urlprefix\url{https://www.sciencedirect.com/science/article/pii/S0022459616302997}.

\bibitem[{\citenamefont{Li et~al.}(2022)\citenamefont{Li, Gu, Chen, Garlea,
  Iida, Kamazawa, Li, Deng, Xiao, Zheng et~al.}}]{Li}
\bibinfo{author}{\bibfnamefont{X.}~\bibnamefont{Li}},
  \bibinfo{author}{\bibfnamefont{Y.}~\bibnamefont{Gu}},
  \bibinfo{author}{\bibfnamefont{Y.}~\bibnamefont{Chen}},
  \bibinfo{author}{\bibfnamefont{V.~O.} \bibnamefont{Garlea}},
  \bibinfo{author}{\bibfnamefont{K.}~\bibnamefont{Iida}},
  \bibinfo{author}{\bibfnamefont{K.}~\bibnamefont{Kamazawa}},
  \bibinfo{author}{\bibfnamefont{Y.}~\bibnamefont{Li}},
  \bibinfo{author}{\bibfnamefont{G.}~\bibnamefont{Deng}},
  \bibinfo{author}{\bibfnamefont{Q.}~\bibnamefont{Xiao}},
  \bibinfo{author}{\bibfnamefont{X.}~\bibnamefont{Zheng}},
  \bibnamefont{et~al.}, \emph{\bibinfo{title}{Giant magnetic in-plane
  anisotropy and competing instabilities in {N}a$_3${C}o$_2${S}b{O}$_6$}}
  (\bibinfo{year}{2022}), \eprint{2204.04593}.

\bibitem[{\citenamefont{Sanders et~al.}(2022)\citenamefont{Sanders, Mole, Liu,
  Brown, Yu, Ling, and Rachel}}]{Sanders2022}
\bibinfo{author}{\bibfnamefont{A.~L.} \bibnamefont{Sanders}},
  \bibinfo{author}{\bibfnamefont{R.~A.} \bibnamefont{Mole}},
  \bibinfo{author}{\bibfnamefont{J.}~\bibnamefont{Liu}},
  \bibinfo{author}{\bibfnamefont{A.~J.} \bibnamefont{Brown}},
  \bibinfo{author}{\bibfnamefont{D.}~\bibnamefont{Yu}},
  \bibinfo{author}{\bibfnamefont{C.~D.} \bibnamefont{Ling}}, \bibnamefont{and}
  \bibinfo{author}{\bibfnamefont{S.}~\bibnamefont{Rachel}},
  \bibinfo{journal}{Physical Review B} \textbf{\bibinfo{volume}{106}},
  \bibinfo{pages}{014413} (\bibinfo{year}{2022}),
  \urlprefix\url{http://arxiv.org/abs/2112.12254}.

\bibitem[{\citenamefont{Songvilay et~al.}(2020)\citenamefont{Songvilay, Robert,
  Petit, Rodriguez-Rivera, Ratcliff, Damay, Bal{\'e}dent, Jim{\'e}nez-Ruiz,
  Lejay, Pachoud et~al.}}]{songvilay2020}
\bibinfo{author}{\bibfnamefont{M.}~\bibnamefont{Songvilay}},
  \bibinfo{author}{\bibfnamefont{J.}~\bibnamefont{Robert}},
  \bibinfo{author}{\bibfnamefont{S.}~\bibnamefont{Petit}},
  \bibinfo{author}{\bibfnamefont{J.}~\bibnamefont{Rodriguez-Rivera}},
  \bibinfo{author}{\bibfnamefont{W.}~\bibnamefont{Ratcliff}},
  \bibinfo{author}{\bibfnamefont{F.}~\bibnamefont{Damay}},
  \bibinfo{author}{\bibfnamefont{V.}~\bibnamefont{Bal{\'e}dent}},
  \bibinfo{author}{\bibfnamefont{M.}~\bibnamefont{Jim{\'e}nez-Ruiz}},
  \bibinfo{author}{\bibfnamefont{P.}~\bibnamefont{Lejay}},
  \bibinfo{author}{\bibfnamefont{E.}~\bibnamefont{Pachoud}},
  \bibnamefont{et~al.}, \bibinfo{journal}{Physical Review B}
  \textbf{\bibinfo{volume}{102}}, \bibinfo{pages}{224429}
  (\bibinfo{year}{2020}).

\bibitem[{\citenamefont{Kim et~al.}(2021)\citenamefont{Kim, Jeong, Lin, Park,
  Masuda, Asai, Itoh, Kim, Zhou, Ma et~al.}}]{Kim}
\bibinfo{author}{\bibfnamefont{C.}~\bibnamefont{Kim}},
  \bibinfo{author}{\bibfnamefont{J.}~\bibnamefont{Jeong}},
  \bibinfo{author}{\bibfnamefont{G.}~\bibnamefont{Lin}},
  \bibinfo{author}{\bibfnamefont{P.}~\bibnamefont{Park}},
  \bibinfo{author}{\bibfnamefont{T.}~\bibnamefont{Masuda}},
  \bibinfo{author}{\bibfnamefont{S.}~\bibnamefont{Asai}},
  \bibinfo{author}{\bibfnamefont{S.}~\bibnamefont{Itoh}},
  \bibinfo{author}{\bibfnamefont{H.-S.} \bibnamefont{Kim}},
  \bibinfo{author}{\bibfnamefont{H.}~\bibnamefont{Zhou}},
  \bibinfo{author}{\bibfnamefont{J.}~\bibnamefont{Ma}}, \bibnamefont{et~al.},
  \bibinfo{journal}{Journal of Physics: Condensed Matter}
  \textbf{\bibinfo{volume}{34}}, \bibinfo{pages}{045802}
  (\bibinfo{year}{2021}),
  \urlprefix\url{https://doi.org/10.1088/1361-648x/ac2644}.

\bibitem[{\citenamefont{Kikuchi et~al.}()\citenamefont{Kikuchi, Kamoda, Mera,
  Takahashi, Okumura, and Yasui}}]{Kikuchi}
\bibinfo{author}{\bibfnamefont{J.}~\bibnamefont{Kikuchi}},
  \bibinfo{author}{\bibfnamefont{T.}~\bibnamefont{Kamoda}},
  \bibinfo{author}{\bibfnamefont{N.}~\bibnamefont{Mera}},
  \bibinfo{author}{\bibfnamefont{Y.}~\bibnamefont{Takahashi}},
  \bibinfo{author}{\bibfnamefont{K.}~\bibnamefont{Okumura}}, \bibnamefont{and}
  \bibinfo{author}{\bibfnamefont{Y.}~\bibnamefont{Yasui}},
  \bibinfo{journal}{arXiv.2206.05409}  (????).

\bibitem[{\citenamefont{Stratan et~al.}(2019)\citenamefont{Stratan, Shukaev,
  Vasilchikova, Vasiliev, Korshunov, Kurbakov, Nalbandyan, and
  Zvereva}}]{Stratan}
\bibinfo{author}{\bibfnamefont{M.~I.} \bibnamefont{Stratan}},
  \bibinfo{author}{\bibfnamefont{I.~L.} \bibnamefont{Shukaev}},
  \bibinfo{author}{\bibfnamefont{T.~M.} \bibnamefont{Vasilchikova}},
  \bibinfo{author}{\bibfnamefont{A.~N.} \bibnamefont{Vasiliev}},
  \bibinfo{author}{\bibfnamefont{A.~N.} \bibnamefont{Korshunov}},
  \bibinfo{author}{\bibfnamefont{A.~I.} \bibnamefont{Kurbakov}},
  \bibinfo{author}{\bibfnamefont{V.~B.} \bibnamefont{Nalbandyan}},
  \bibnamefont{and} \bibinfo{author}{\bibfnamefont{E.~A.}
  \bibnamefont{Zvereva}}, \bibinfo{journal}{New J. Chem.}
  \textbf{\bibinfo{volume}{43}}, \bibinfo{pages}{13545} (\bibinfo{year}{2019}),
  \urlprefix\url{http://dx.doi.org/10.1039/C9NJ03627J}.

\bibitem[{\citenamefont{Roisnel and Rodr{\'{\i}}quez-Carvajal}(2001)}]{Roisnel}
\bibinfo{author}{\bibfnamefont{T.}~\bibnamefont{Roisnel}} \bibnamefont{and}
  \bibinfo{author}{\bibfnamefont{J.}~\bibnamefont{Rodr{\'{\i}}quez-Carvajal}},
  in \emph{\bibinfo{booktitle}{European Powder Diffraction EPDIC 7}}
  (\bibinfo{publisher}{Trans Tech Publications Ltd}, \bibinfo{year}{2001}),
  vol. \bibinfo{volume}{378} of \emph{\bibinfo{series}{Materials Science
  Forum}}, pp. \bibinfo{pages}{118--123}.

\bibitem[{\citenamefont{Narath}(1967)}]{Narath}
\bibinfo{author}{\bibfnamefont{A.}~\bibnamefont{Narath}},
  \bibinfo{journal}{Phys. Rev.} \textbf{\bibinfo{volume}{162}},
  \bibinfo{pages}{320} (\bibinfo{year}{1967}),
  \urlprefix\url{https://link.aps.org/doi/10.1103/PhysRev.162.320}.

\bibitem[{\citenamefont{Tari}(2003)}]{Tari}
\bibinfo{author}{\bibfnamefont{A.}~\bibnamefont{Tari}},
  \emph{\bibinfo{title}{The Specific Heat of Matter at Low Temperature}}
  (\bibinfo{publisher}{Imperial College Press, London}, \bibinfo{year}{2003}).

\bibitem[{\citenamefont{de~Jongh and Miedema}(1974)}]{deJongh}
\bibinfo{author}{\bibfnamefont{L.}~\bibnamefont{de~Jongh}} \bibnamefont{and}
  \bibinfo{author}{\bibfnamefont{A.}~\bibnamefont{Miedema}},
  \bibinfo{journal}{Advances in Physics} \textbf{\bibinfo{volume}{23}},
  \bibinfo{pages}{1} (\bibinfo{year}{1974}),
  \urlprefix\url{https://doi.org/10.1080/00018739700101558}.

\bibitem[{\citenamefont{Wolter et~al.}(2017)\citenamefont{Wolter, Corredor,
  Janssen, Nenkov, Sch\"onecker, Do, Choi, Albrecht, Hunger, Doert
  et~al.}}]{Wolter}
\bibinfo{author}{\bibfnamefont{A.~U.~B.} \bibnamefont{Wolter}},
  \bibinfo{author}{\bibfnamefont{L.~T.} \bibnamefont{Corredor}},
  \bibinfo{author}{\bibfnamefont{L.}~\bibnamefont{Janssen}},
  \bibinfo{author}{\bibfnamefont{K.}~\bibnamefont{Nenkov}},
  \bibinfo{author}{\bibfnamefont{S.}~\bibnamefont{Sch\"onecker}},
  \bibinfo{author}{\bibfnamefont{S.-H.} \bibnamefont{Do}},
  \bibinfo{author}{\bibfnamefont{K.-Y.} \bibnamefont{Choi}},
  \bibinfo{author}{\bibfnamefont{R.}~\bibnamefont{Albrecht}},
  \bibinfo{author}{\bibfnamefont{J.}~\bibnamefont{Hunger}},
  \bibinfo{author}{\bibfnamefont{T.}~\bibnamefont{Doert}},
  \bibnamefont{et~al.}, \bibinfo{journal}{Phys. Rev. B}
  \textbf{\bibinfo{volume}{96}}, \bibinfo{pages}{041405}
  (\bibinfo{year}{2017}),
  \urlprefix\url{https://link.aps.org/doi/10.1103/PhysRevB.96.041405}.

\bibitem[{\citenamefont{Sears et~al.}(2017)\citenamefont{Sears, Zhao, Xu, Lynn,
  and Kim}}]{Sears}
\bibinfo{author}{\bibfnamefont{J.~A.} \bibnamefont{Sears}},
  \bibinfo{author}{\bibfnamefont{Y.}~\bibnamefont{Zhao}},
  \bibinfo{author}{\bibfnamefont{Z.}~\bibnamefont{Xu}},
  \bibinfo{author}{\bibfnamefont{J.~W.} \bibnamefont{Lynn}}, \bibnamefont{and}
  \bibinfo{author}{\bibfnamefont{Y.-J.} \bibnamefont{Kim}},
  \bibinfo{journal}{Phys. Rev. B} \textbf{\bibinfo{volume}{95}},
  \bibinfo{pages}{180411} (\bibinfo{year}{2017}),
  \urlprefix\url{https://link.aps.org/doi/10.1103/PhysRevB.95.180411}.

\bibitem[{\citenamefont{Yamada and Sakata}(1986)}]{Yamada}
\bibinfo{author}{\bibfnamefont{Y.}~\bibnamefont{Yamada}} \bibnamefont{and}
  \bibinfo{author}{\bibfnamefont{A.}~\bibnamefont{Sakata}},
  \bibinfo{journal}{Journal of the Physical Society of Japan}
  \textbf{\bibinfo{volume}{55}}, \bibinfo{pages}{1751} (\bibinfo{year}{1986}),
  \eprint{https://doi.org/10.1143/JPSJ.55.1751},
  \urlprefix\url{https://doi.org/10.1143/JPSJ.55.1751}.

\bibitem[{\citenamefont{Borsa et~al.}(1992)\citenamefont{Borsa, Corti, Goto,
  Rigamonti, Johnston, and Chou}}]{Borsa}
\bibinfo{author}{\bibfnamefont{F.}~\bibnamefont{Borsa}},
  \bibinfo{author}{\bibfnamefont{M.}~\bibnamefont{Corti}},
  \bibinfo{author}{\bibfnamefont{T.}~\bibnamefont{Goto}},
  \bibinfo{author}{\bibfnamefont{A.}~\bibnamefont{Rigamonti}},
  \bibinfo{author}{\bibfnamefont{D.~C.} \bibnamefont{Johnston}},
  \bibnamefont{and} \bibinfo{author}{\bibfnamefont{F.~C.} \bibnamefont{Chou}},
  \bibinfo{journal}{Phys. Rev. B} \textbf{\bibinfo{volume}{45}},
  \bibinfo{pages}{5756} (\bibinfo{year}{1992}),
  \urlprefix\url{https://link.aps.org/doi/10.1103/PhysRevB.45.5756}.

\bibitem[{\citenamefont{Kurbakov et~al.}(2022)\citenamefont{Kurbakov,
  Susloparova, Pomjakushin, Skourski, Vavilova, Vasilchikova, Raganyan, and
  Vasiliev}}]{Kurbakov}
\bibinfo{author}{\bibfnamefont{A.~I.} \bibnamefont{Kurbakov}},
  \bibinfo{author}{\bibfnamefont{A.~E.} \bibnamefont{Susloparova}},
  \bibinfo{author}{\bibfnamefont{V.~Y.} \bibnamefont{Pomjakushin}},
  \bibinfo{author}{\bibfnamefont{Y.}~\bibnamefont{Skourski}},
  \bibinfo{author}{\bibfnamefont{E.~L.} \bibnamefont{Vavilova}},
  \bibinfo{author}{\bibfnamefont{T.~M.} \bibnamefont{Vasilchikova}},
  \bibinfo{author}{\bibfnamefont{G.~V.} \bibnamefont{Raganyan}},
  \bibnamefont{and} \bibinfo{author}{\bibfnamefont{A.~N.}
  \bibnamefont{Vasiliev}}, \bibinfo{journal}{Phys. Rev. B}
  \textbf{\bibinfo{volume}{105}}, \bibinfo{pages}{064416}
  (\bibinfo{year}{2022}),
  \urlprefix\url{https://link.aps.org/doi/10.1103/PhysRevB.105.064416}.

\bibitem[{\citenamefont{Itoh}(2015)}]{Itoh}
\bibinfo{author}{\bibfnamefont{Y.}~\bibnamefont{Itoh}},
  \bibinfo{journal}{Journal of the Physical Society of Japan}
  \textbf{\bibinfo{volume}{84}}, \bibinfo{pages}{064714}
  (\bibinfo{year}{2015}),
  \urlprefix\url{https://doi.org/10.7566/JPSJ.84.064714}.

\bibitem[{\citenamefont{Plumer and Mailhot}(1994)}]{Plumer}
\bibinfo{author}{\bibfnamefont{M.~L.} \bibnamefont{Plumer}} \bibnamefont{and}
  \bibinfo{author}{\bibfnamefont{A.}~\bibnamefont{Mailhot}},
  \bibinfo{journal}{Phys. Rev. B} \textbf{\bibinfo{volume}{50}},
  \bibinfo{pages}{16113} (\bibinfo{year}{1994}),
  \urlprefix\url{https://link.aps.org/doi/10.1103/PhysRevB.50.16113}.

\bibitem[{\citenamefont{Zhang and Yang}(1994)}]{Zhang}
\bibinfo{author}{\bibfnamefont{G.~M.} \bibnamefont{Zhang}} \bibnamefont{and}
  \bibinfo{author}{\bibfnamefont{C.~Z.} \bibnamefont{Yang}},
  \bibinfo{journal}{Phys. Rev. B} \textbf{\bibinfo{volume}{50}},
  \bibinfo{pages}{12546} (\bibinfo{year}{1994}),
  \urlprefix\url{https://link.aps.org/doi/10.1103/PhysRevB.50.12546}.

\bibitem[{\citenamefont{Slichter}(1989)}]{Slichter}
\bibinfo{author}{\bibfnamefont{C.}~\bibnamefont{Slichter}},
  \emph{\bibinfo{title}{Principles of Magnetic Resonance}}
  (\bibinfo{publisher}{Springer, New York}, \bibinfo{year}{1989}).

\bibitem[{\citenamefont{Narath and Fromhold}(1966)}]{Narath2}
\bibinfo{author}{\bibfnamefont{A.}~\bibnamefont{Narath}} \bibnamefont{and}
  \bibinfo{author}{\bibfnamefont{A.~T.} \bibnamefont{Fromhold}},
  \bibinfo{journal}{Phys. Rev. Lett.} \textbf{\bibinfo{volume}{17}},
  \bibinfo{pages}{354} (\bibinfo{year}{1966}),
  \urlprefix\url{https://link.aps.org/doi/10.1103/PhysRevLett.17.354}.

\bibitem[{\citenamefont{Janša et~al.}(2018)\citenamefont{Janša, Zorko,
  Gomilšek, Pregelj, Krämer, Biner, Biffin, Rüegg, and Klanjšek}}]{Jansa}
\bibinfo{author}{\bibfnamefont{N.}~\bibnamefont{Janša}},
  \bibinfo{author}{\bibfnamefont{A.}~\bibnamefont{Zorko}},
  \bibinfo{author}{\bibfnamefont{M.}~\bibnamefont{Gomilšek}},
  \bibinfo{author}{\bibfnamefont{M.}~\bibnamefont{Pregelj}},
  \bibinfo{author}{\bibfnamefont{K.~W.} \bibnamefont{Krämer}},
  \bibinfo{author}{\bibfnamefont{D.}~\bibnamefont{Biner}},
  \bibinfo{author}{\bibfnamefont{A.}~\bibnamefont{Biffin}},
  \bibinfo{author}{\bibfnamefont{C.}~\bibnamefont{Rüegg}}, \bibnamefont{and}
  \bibinfo{author}{\bibfnamefont{M.}~\bibnamefont{Klanjšek}},
  \bibinfo{journal}{Nature Physics} \textbf{\bibinfo{volume}{14}},
  \bibinfo{pages}{786–790} (\bibinfo{year}{2018}),
  \urlprefix\url{https://www.nature.com/articles/s41567-018-0129-5}.

\bibitem[{\citenamefont{Nagai et~al.}(2020)\citenamefont{Nagai, Jinno,
  Yoshitake, Nasu, Motome, Itoh, and Shimizu}}]{Nagai}
\bibinfo{author}{\bibfnamefont{Y.}~\bibnamefont{Nagai}},
  \bibinfo{author}{\bibfnamefont{T.}~\bibnamefont{Jinno}},
  \bibinfo{author}{\bibfnamefont{J.}~\bibnamefont{Yoshitake}},
  \bibinfo{author}{\bibfnamefont{J.}~\bibnamefont{Nasu}},
  \bibinfo{author}{\bibfnamefont{Y.}~\bibnamefont{Motome}},
  \bibinfo{author}{\bibfnamefont{M.}~\bibnamefont{Itoh}}, \bibnamefont{and}
  \bibinfo{author}{\bibfnamefont{Y.}~\bibnamefont{Shimizu}},
  \bibinfo{journal}{Phys. Rev. B} \textbf{\bibinfo{volume}{101}},
  \bibinfo{pages}{020414} (\bibinfo{year}{2020}),
  \urlprefix\url{https://link.aps.org/doi/10.1103/PhysRevB.101.020414}.

\bibitem[{\citenamefont{Yoshitake et~al.}(2016)\citenamefont{Yoshitake, Nasu,
  and Motome}}]{Yoshitake}
\bibinfo{author}{\bibfnamefont{J.}~\bibnamefont{Yoshitake}},
  \bibinfo{author}{\bibfnamefont{J.}~\bibnamefont{Nasu}}, \bibnamefont{and}
  \bibinfo{author}{\bibfnamefont{Y.}~\bibnamefont{Motome}},
  \bibinfo{journal}{Phys. Rev. Lett.} \textbf{\bibinfo{volume}{117}},
  \bibinfo{pages}{157203} (\bibinfo{year}{2016}),
  \urlprefix\url{https://link.aps.org/doi/10.1103/PhysRevLett.117.157203}.

\bibitem[{\citenamefont{Kitaev}(2006)}]{Kitaev}
\bibinfo{author}{\bibfnamefont{A.}~\bibnamefont{Kitaev}},
  \bibinfo{journal}{Annals of Physics} \textbf{\bibinfo{volume}{321}},
  \bibinfo{pages}{2} (\bibinfo{year}{2006}), ISSN \bibinfo{issn}{00034916},
  \urlprefix\url{http://linkinghub.elsevier.com/retrieve/pii/S0003491605002381}.

\bibitem[{\citenamefont{Song et~al.}(2016)\citenamefont{Song, You, and
  Balents}}]{Song}
\bibinfo{author}{\bibfnamefont{X.-Y.} \bibnamefont{Song}},
  \bibinfo{author}{\bibfnamefont{Y.-Z.} \bibnamefont{You}}, \bibnamefont{and}
  \bibinfo{author}{\bibfnamefont{L.}~\bibnamefont{Balents}},
  \bibinfo{journal}{Phys. Rev. Lett.} \textbf{\bibinfo{volume}{117}},
  \bibinfo{pages}{037209} (\bibinfo{year}{2016}),
  \urlprefix\url{https://link.aps.org/doi/10.1103/PhysRevLett.117.037209}.

\bibitem[{\citenamefont{Wu et~al.}(2022)\citenamefont{Wu, Katsura, Li, Cai, and
  Guan}}]{Wu}
\bibinfo{author}{\bibfnamefont{N.}~\bibnamefont{Wu}},
  \bibinfo{author}{\bibfnamefont{H.}~\bibnamefont{Katsura}},
  \bibinfo{author}{\bibfnamefont{S.-W.} \bibnamefont{Li}},
  \bibinfo{author}{\bibfnamefont{X.}~\bibnamefont{Cai}}, \bibnamefont{and}
  \bibinfo{author}{\bibfnamefont{X.-W.} \bibnamefont{Guan}},
  \bibinfo{journal}{Phys. Rev. B} \textbf{\bibinfo{volume}{105}},
  \bibinfo{pages}{064419} (\bibinfo{year}{2022}),
  \urlprefix\url{https://link.aps.org/doi/10.1103/PhysRevB.105.064419}.

\bibitem[{\citenamefont{Li et~al.}(2021)\citenamefont{Li, Zhang, Wang, Wu, Gao,
  Qu, Liu, Gong, and Li}}]{Li2021}
\bibinfo{author}{\bibfnamefont{H.}~\bibnamefont{Li}},
  \bibinfo{author}{\bibfnamefont{H.-K.} \bibnamefont{Zhang}},
  \bibinfo{author}{\bibfnamefont{J.}~\bibnamefont{Wang}},
  \bibinfo{author}{\bibfnamefont{H.-Q.} \bibnamefont{Wu}},
  \bibinfo{author}{\bibfnamefont{Y.}~\bibnamefont{Gao}},
  \bibinfo{author}{\bibfnamefont{D.-W.} \bibnamefont{Qu}},
  \bibinfo{author}{\bibfnamefont{Z.-X.} \bibnamefont{Liu}},
  \bibinfo{author}{\bibfnamefont{S.-S.} \bibnamefont{Gong}}, \bibnamefont{and}
  \bibinfo{author}{\bibfnamefont{W.}~\bibnamefont{Li}},
  \bibinfo{journal}{Nature Communications} \textbf{\bibinfo{volume}{12}},
  \bibinfo{pages}{4007} (\bibinfo{year}{2021}).

\bibitem[{\citenamefont{Lee et~al.}(2021)\citenamefont{Lee, Lee, Choi, Jang,
  Kalaivanan, Sankar, and Choi}}]{Choi}
\bibinfo{author}{\bibfnamefont{C.~H.} \bibnamefont{Lee}},
  \bibinfo{author}{\bibfnamefont{S.}~\bibnamefont{Lee}},
  \bibinfo{author}{\bibfnamefont{Y.~S.} \bibnamefont{Choi}},
  \bibinfo{author}{\bibfnamefont{Z.~H.} \bibnamefont{Jang}},
  \bibinfo{author}{\bibfnamefont{R.}~\bibnamefont{Kalaivanan}},
  \bibinfo{author}{\bibfnamefont{R.}~\bibnamefont{Sankar}}, \bibnamefont{and}
  \bibinfo{author}{\bibfnamefont{K.-Y.} \bibnamefont{Choi}},
  \bibinfo{journal}{Phys. Rev. B} \textbf{\bibinfo{volume}{103}},
  \bibinfo{pages}{214447} (\bibinfo{year}{2021}),
  \urlprefix\url{https://link.aps.org/doi/10.1103/PhysRevB.103.214447}.

\bibitem[{\citenamefont{Kuo et~al.}(2012)\citenamefont{Kuo, Jian, and
  Lue}}]{Kuo}
\bibinfo{author}{\bibfnamefont{C.}~\bibnamefont{Kuo}},
  \bibinfo{author}{\bibfnamefont{T.}~\bibnamefont{Jian}}, \bibnamefont{and}
  \bibinfo{author}{\bibfnamefont{C.}~\bibnamefont{Lue}},
  \bibinfo{journal}{Journal of Alloys and Compounds}
  \textbf{\bibinfo{volume}{531}}, \bibinfo{pages}{1} (\bibinfo{year}{2012}),
  ISSN \bibinfo{issn}{0925-8388},
  \urlprefix\url{https://www.sciencedirect.com/science/article/pii/S0925838812004100}.

\bibitem[{\citenamefont{Berthelot et~al.}(2012)\citenamefont{Berthelot,
  Schmidt, Sleight, and Subramanian}}]{Berthelot}
\bibinfo{author}{\bibfnamefont{R.}~\bibnamefont{Berthelot}},
  \bibinfo{author}{\bibfnamefont{W.}~\bibnamefont{Schmidt}},
  \bibinfo{author}{\bibfnamefont{A.}~\bibnamefont{Sleight}}, \bibnamefont{and}
  \bibinfo{author}{\bibfnamefont{M.}~\bibnamefont{Subramanian}},
  \bibinfo{journal}{Journal of Solid State Chemistry}
  \textbf{\bibinfo{volume}{196}}, \bibinfo{pages}{225} (\bibinfo{year}{2012}),
  ISSN \bibinfo{issn}{0022-4596},
  \urlprefix\url{https://www.sciencedirect.com/science/article/pii/S0022459612003970}.

\bibitem[{\citenamefont{Moriya}(1963)}]{Moriya}
\bibinfo{author}{\bibfnamefont{T.}~\bibnamefont{Moriya}},
  \bibinfo{journal}{Journal of the Physical Society of Japan}
  \textbf{\bibinfo{volume}{18}}, \bibinfo{pages}{516} (\bibinfo{year}{1963}),
  \eprint{https://doi.org/10.1143/JPSJ.18.516},
  \urlprefix\url{https://doi.org/10.1143/JPSJ.18.516}.

\bibitem[{\citenamefont{Bałanda}(2013)}]{Balanda}
\bibinfo{author}{\bibfnamefont{M.}~\bibnamefont{Bałanda}},
  \bibinfo{journal}{ACTA PHYSICA POLONICA A} \textbf{\bibinfo{volume}{124}},
  \bibinfo{pages}{964} (\bibinfo{year}{2013}),
  \urlprefix\url{10.12693/APhysPolA.124.964}.

\end{thebibliography}

\end{document}